\documentclass[twocolumn]{aastex61}
\usepackage{bm}
\usepackage{amsmath}
\usepackage{booktabs}
\usepackage{graphicx}
\usepackage{comment}

\newcommand{\given}{\,|\,}
\newcommand{\dd}{\mathrm{d}}

\newcommand{\degree}{\ifmmode {^{\circ}\ }\else$^{\circ}$\fi}
\newcommand{\amin}{\ifmmode {^{\prime}\ }\else$^{\prime}$\fi}
\newcommand{\asec}{\ifmmode {^{\prime\prime}}\else$^{\prime\prime}$\fi}

\newcommand{\ergs}{\ifmmode { {\rm erg}\ {\rm s}^{-1} } \else erg s$^{-1}$\fi}

\newcommand{\Msun}{\ifmmode {M_{\odot}}\else${M_{\odot}}$\fi}
\newcommand{\Rsun}{\ifmmode {R_{\odot}}\else${R_{\odot}}$\fi}
\newcommand{\Porb}{\ifmmode {P_{\rm orb}}\else${P_{\rm orb}}$\fi}

\newcommand{\bse}{{\tt BSE}}
\newcommand{\dart}{{\tt dart\_board}}
\newcommand{\emcee}{{\tt emcee}}

\submitjournal{ApJS}

\shorttitle{\dart: pop.\ synth.\ with MCMC}
\shortauthors{Andrews et al.}

\begin{document}

\title{\dart: Binary Population Synthesis with Markov Chain Monte Carlo}

\correspondingauthor{Jeff J.\ Andrews}
\email{andrews@physics.uoc.gr}

\author[0000-0001-5261-3923]{Jeff J.\ Andrews}
\affiliation{Foundation for Research and Technology-Hellas, 
100 Nikolaou Plastira St., 
71110 Heraklion, Crete, Greece}
\affiliation{Physics Department \& Institute of Theoretical \& Computational Physics, 
P.O Box 2208, 
71003 Heraklion, Crete, Greece}

\author{Andreas Zezas}
\affiliation{Foundation for Research and Technology-Hellas, 
100 Nikolaou Plastira St., 
71110 Heraklion, Crete, Greece}
\affiliation{Physics Department \& Institute of Theoretical \& Computational Physics, 
P.O Box 2208, 
71003 Heraklion, Crete, Greece}
\affiliation{Harvard-Smithsonian Center for Astrophysics,  
60 Garden Street, 
Cambridge, MA 02138, USA}

\author[0000-0003-1474-1523]{Tassos Fragos}
\affiliation{Geneva Observatory, 
University of Geneva, 
Chemin des Maillettes 51, 
1290 Sauverny, Switzerland}
\affiliation{DARK, Niels Bohr Institute, 
University of Copenhagen, 
 Juliane Maries Vej 30, 
 DK-2100 Copenhagen, Denmark}

\begin{abstract}
By employing Monte Carlo random sampling, traditional binary population synthesis (BPS) offers a substantial improvement in efficiency over brute force, grid-based studies. Even so, BPS models typically require a large number of simulation realizations, a computationally expensive endeavor, to generate statistically robust results. Recent advances in statistical methods have led us to revisit the traditional approach to BPS. In this work we describe our publicly available code \dart\ which combines rapid binary evolution codes, typically used in traditional BPS, with modern Markov chain Monte Carlo methods. \dart\ takes a novel approach that treats the initial binary parameters and the supernova kick vector as model parameters. This formulation has several advantages, including the ability to model either populations of systems or individual binaries, the natural inclusion of observational uncertainties, and the flexible addition of new constraints which are problematic to include using traditional BPS. After testing our code with mock systems, we demonstrate the flexibility of \dart\ by applying it to three examples: (i) a generic population of high mass X-ray binaries (HMXBs), (ii) the population of HMXBs in the Large Magellanic Cloud (LMC) in which the spatially resolved star formation history is used as a prior, and (iii) one particular HMXB in the LMC, {\it Swift} J0513.4$-$6547, in which we include observations of the system's component masses and orbital period. Although this work focuses on HMXBs, \dart\ can be applied to a variety of stellar binaries including the recent detections by gravitational wave observatories of merging compact object binaries.
\end{abstract}


\section{Introduction}
\label{sec:intro}

Stellar binaries have been shown to affect a broad range of astrophysical systems, ranging from the prevalence of exoplanets \citep{kraus16} to the cumulative effect of interacting binaries in unresolved stellar populations \citep{eldridge17}. Recent estimates indicate that roughly 70\% of all massive binaries interact \citep{sana12}, leading to a variety of astrophysical exotica including Type Ibc SNe \citep{smith11}, luminous blue variables \citep{smith15}, and the recent detection of gravitational waves from binary black holes \citep{abbott16a} and binary neutron stars \citep{ligo17b}. For a recent review on the impact of stellar binaries, see \citet{demarco17}.

Efforts to theoretically model populations of binary stars have traditionally relied on binary population synthesis (BPS) in which one randomly generates, according to some predetermined initial probability distributions, a large number of stellar binaries \citep[e.g.,][]{portegieszwart96, lipunov97, tout97}. Using our knowledge of astrophysics built into a rapid binary evolution code, these synthetic binaries are evolved until their present state, when one takes a ``snapshot'' of the modeled systems \citep[for a recent discussion of state-of-the-art BPS codes and their differences, see][]{toonen14}. The resulting samples are used to understand the evolutionary history of individual systems \citep[e.g.,][]{sorensen17}, make predictions (such as rate estimates) for future observations \citep[e.g.,][]{ablimit16}, and constrain binary evolution physics by comparing the set of simulated systems to a well-characterized observational sample \citep[e.g.,][]{andrews15}.

Despite its common use, BPS can be an inefficient tool for rare or short-lived evolutionary states since significant computational time is spent on regions of parameter space of no interest to the observed systems. For instance, when studying neutron star (NS) or black hole (BH) binaries, many or even most simulated systems disrupt due to the natal kick imparted to the compact object during the supernova (SN). Other astrophysical processes such as the unstable mass transfer phase, known as a common envelope, may cause a large fraction of simulated systems to merge \citep[for a recent review, see][]{ivanova13}. Unfortunately, there is no way to know {\it a priori} which set of initial binary conditions form the systems of interest; the entire region of plausible parameter space must be tested. Therefore, in BPS studies, more than $10^6$ binaries are often required to make even qualitative comparisons with observed populations. For instance, in a recent study of the ultraluminous X-ray binary (ULX) M82 X-2, \citet{fragos15} generated $10^7$ initial binaries, finding only $10^2-10^3$ evolved into systems matching the observational characteristics of M82 X-2. An alternative to traditional BPS is desirable\footnote{For a small enough parameter space and substantial computing power, grid-based studies may be sufficient \citep[e.g.,][]{marchant17}.}.

One approach uses a Jacobian formalism to transform initial binary probability distributions to distributions of observed parameters. This method has been developed by \citet{kolb93} and \citet{politano96} for cataclysmic variables and extended by \citet{kalogera96} for high mass binaries, including SN kicks \citep[see also][]{kalogera98, kalogera00}. Jacobian transformations have most recently been employed by \citet{bhadkamkar12,bhadkamkar14} to describe populations of high mass X-ray binaries (HMXB) and low mass X-ray binaries (LMXB), respectively.\footnote{X-ray binaries are comprised of an NS or BH accreting mass from a non-degenerate companion star of either a low or high mass, depending on whether it is an LMXB or HMXB, respectively. The exact mass separating an LMXB from an HMXB is somewhat arbitrary; throughout this work, we define HMXBs as having donor stars with masses above 6 \Msun.} Although analytic and efficient, these works lack flexibility and can only approximate key binary evolution physics. It remains to be seen whether analytic methods can incorporate the level of detail required to provide more than qualitative comparisons with observations.

In this work, we describe \dart, an open-source code written in {\tt python} that provides a statistical wrapper to rapid binary evolution codes. We consider the initial binary parameters (including the SN kick magnitude and direction) as model parameters with prior probabilities based on the same initial distributions used by traditional BPS. Our likelihood function flexibly combines available binary observables, which allows our method to be adaptable to model either individual systems or a population. We employ a Markov chain Monte Carlo (MCMC) algorithm to search the parameter space of initial binary conditions. Crucially, because MCMC focuses computational power based on the posterior probability (rather than the prior distributions as in traditional BPS), little computational time is wasted on evolving binaries that disrupt or merge. Throughout this work, we use a modified version of the widely used rapid BPS code {\tt BSE} \citep{hurley00,hurley02} with {\tt python} bindings within \dart. However, \dart\ can be easily adapted to be used with any BPS code.

We choose to demonstrate the viability of this method by modeling populations of HMXBs which are comprised of a neutron star or black hole accreting material from an early-type star \citep[for reviews, see][]{ bhattacharya91, tauris06}. This choice is motivated principally by the exquisite quality of the observational sample which has grown immensely over the past decade. With its unprecedented angular resolution, the space-based X-ray observatory {\it Chandra} has identified hundreds of X-ray point sources in nearby galaxies \citep[e.g.,][]{sarazin01,fabbiano01,fabbiano06, wang16}. Studies of these objects, both observational and theoretical using BPS, have yielded a deeper insight into the physical processes forming individual accreting stellar sources, including LMXBs and HMXBs \citep[e.g.,][]{belczynski04, fragos08, fragos09, lehmer10, luo12, tzanavaris13}, as well as ULXs \citep{swartz04,feng11,kaaret17}. The relatively short lifetimes of HMXBs imply that they are indicators of recent star formation, and indeed extragalactic observations find that the contribution from HMXBs to the collective X-ray luminosity of a galaxy increases with increasing star formation rate \citep{grimm03,lehmer10,mineo12}.

The nearby Small Magellanic Cloud (SMC) and Large Magellanic Cloud (LMC) have the best-studied extragalactic X-ray populations. X-ray campaigns with {\it Chandra} \citep[e.g.,][]{laycock10, hong17} and {\it XMM-Newton} \citep[e.g.,][]{sturm13} have brought the number of candidate and confirmed HMXBs in the SMC to 148 \citep[for the most recent catalog, see][]{haberl16}, and ongoing observations are identifying and characterizing new X-ray objects in the somewhat larger LMC \citep{antoniou16, haberl17}. At the same time, infrared, optical, and ultraviolet imaging provide detailed spatially resolved star formation histories (SFH) of regions within the SMC and LMC, with angular resolutions as small as 12\amin\ by 12\amin\ \citep{harris04,harris09}. These SFHs are precise, particularly so in the past 10$^8$ yr when HMXBs were formed.

Observational studies comparing X-ray populations with star forming regions have shown that HMXBs are typically found near regions with recent star formation \citep{zezas02a, kaaret04, antoniou10}. Moreover, \citet{kaaret04} found that more luminous X-ray objects are typically found closer to star forming regions, implying that the kick imparted to the system during the primary star's core collapse was either smaller or only applied recently in the history of the system.

By including information from the SFH, BPS models can add extra constraints to the population of observed HMXBs. \citet{sepinsky05} demonstrated that the distance traveled by a population of HMXBs depends on the binary physics assumed. Later, \citet{zuo10} and \citet{zuo15} went further by correlating this travel distance with orbital period and X-ray luminosity. A more in-depth study of HMXB travel distances, including either the formation of individual systems or spatially resolved SFHs, is lacking; however, it is the next step in modeling X-ray binary populations given the increasing availability of spatially resolved SFHs.

Using the binary's birth position as model parameters and the SFH as a prior on the birth position and time, \dart\ can constrain the evolution of specific systems by comparing a system's current properties and its coordinates on the sky with SFH maps. This increases the constraining power of our method since it incorporates the current positions of binaries in the fit.

\begin{figure}
\begin{center}
\includegraphics[width=0.8\columnwidth]{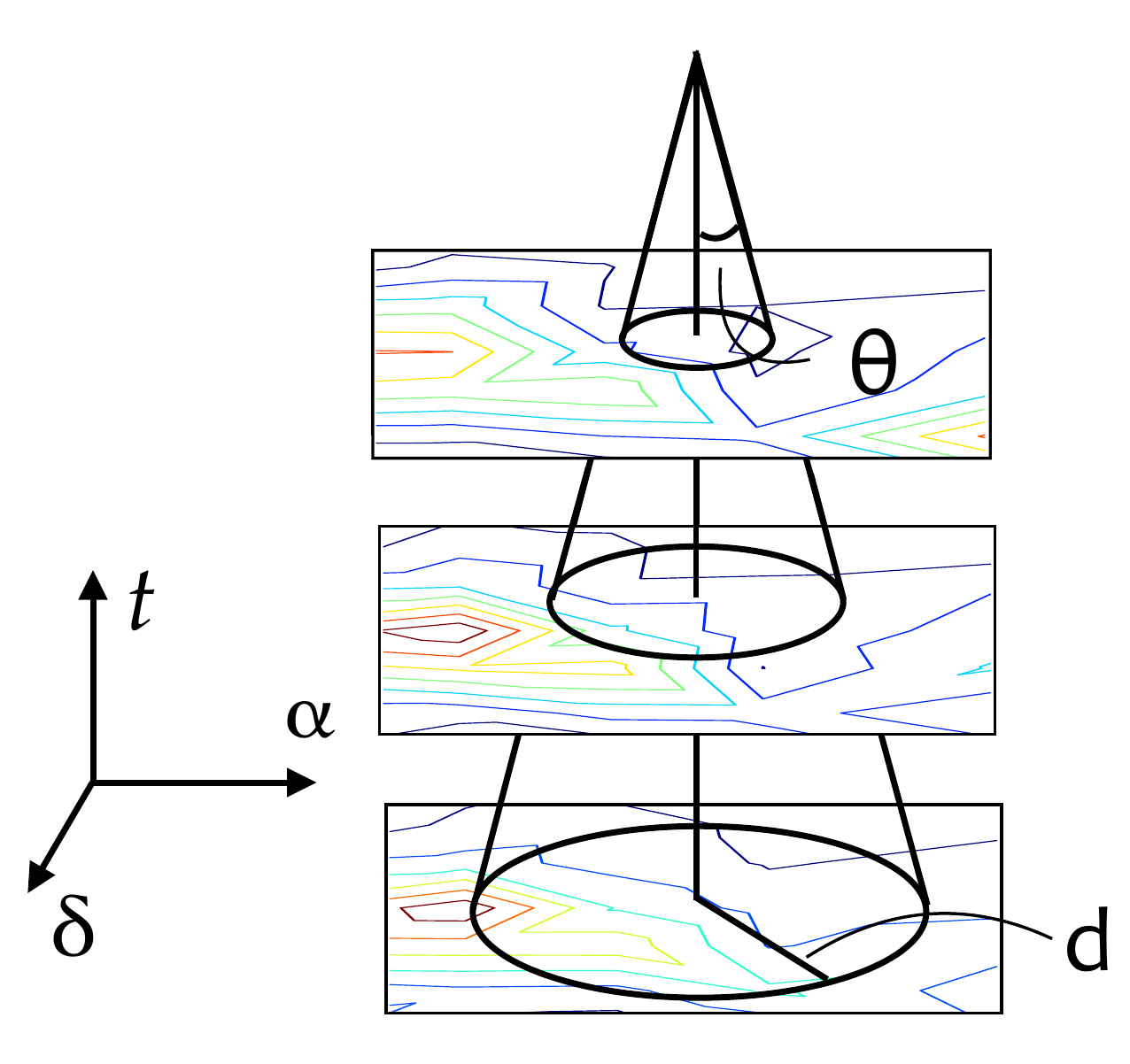}
\caption{ For an observed system today (at the point of the cone), possible birth locations form axisymmetric circles on the sky that increase progressively for longer times since the primary underwent core collapse. The systemic velocity sets the slope of the cone. A putative system at the position shown is likely to have travelled longer, from the region of high star formation indicated by red contours at the bottom left. Therefore, in principle the position of a system near a region of high star formation can constrain the system's age and overall formation. A second region of high star formation exists at the bottom right. However, the system is unlikely to have formed here since, although this region currently has a high star formation rate, the rate was much lower in the past, in the region accessible to this system given its velocity. }
\label{fig:prior_SFH}
\end{center}
\end{figure}

Figure \ref{fig:prior_SFH} shows conceptually how this can be done. Systems that have traveled for a longer time since the primary star's core collapse, could have been formed at an increasingly wider region on the sky. In the example shown, the HMXB may have formed at the peak of high star formation (red contour) to the left of the system (lower right ascension, $\alpha$), but could only have done so if the system has had enough time to travel there. The region of possible formation is defined by a cone with the areas of possible birth positions ($\alpha$, $\delta$) forming concentric circles that become progressively larger as the system travels away from its birth position for longer\footnote{Note that the effect from the gravitational potential of the host galaxy is minimal for typical HMXBs with a lifetime of $\sim$1 Myr and a systemic velocity of dozens of km s$^{-1}$. The velocities of LMXBs, on the other hand, are strongly affected by their host galaxy since these systems have lifetimes of Gyr.}. The shape of the cone is determined by the specific parameters of the binary (the slope is determined by the post-supernova systemic velocity, and the height by the time since the primary's core collapse). Clearly, for a model to make meaningful constraints, the formation of the system must simultaneously account for both the system's evolution and the spatially resolved SFH.

To our knowledge this is the first time that MCMC methods have been combined with BPS. However, many groups have recently developed methods to combine population synthesis with Bayesian inference, including matching stellar populations with isochrones \citep{stenning16}, deriving galactic properties from photometry and spectra \citep{krumholz15}, and obtaining photometric redshifts \citep{tanaka15}, among others. Specific to HMXBs, \citet{douna15} developed an MCMC approach to correlate the X-ray luminosity function from a population of HMXBs with a galaxy's star formation rate and metallicity. Previous works by \citet{ihm06} and \citet{andrews15} each used Bayesian statistical techniques in post-processing to compare traditional BPS results with the sample of known double neutron stars (NS).

In Section \ref{sec:stats} we describe our statistical method and provide the relevant prior and posterior distributions. We test our model on three individual mock systems in Section \ref{sec:mock}. We then apply our model to a general population of HMXBs in Section \ref{sec:results_population}, the population of HMXBs in the LMC in Section \ref{sec:results_population_LMC}, and one specific HMXB in the LMC in Section \ref{sec:J0513}. Finally, we place our method in the broader context of binary population studies, providing some limitations and future directions as well as our conclusions in Section \ref{sec:discussion}.

\section{Statistical Method}
\label{sec:stats}

In this section we define our statistical method, with specific attention to the differences between our method and traditional population synthesis. In Sections \ref{sec:stats_population} and \ref{sec:stats_individual} we demonstrate how our method can be used to model populations of binaries and individual binaries, respectively. Following that, we describe our prior probabilities on model parameters in Section \ref{sec:priors} and the likelihood functions in Section \ref{sec:likelihoods}.

\startlongtable
\begin{deluxetable}{c|c}
\tablecaption{Variable Index \label{tab:variable_index}}
\tablehead{
\colhead{Name} &
\colhead{Description}
}
\startdata
    $\bm{x}_{\rm i}$  &   Set of initial binary parameters \\
    $\bm{x}_{\rm f}$  &   Set of final binary parameters \\
    $M_1$   & Primary mass \\
    $M_2$   & Secondary mass \\
    $a$     & Orbital separation \\
    $P_{\rm orb}$   & Orbital period \\
    $e$     & Orbital eccentricity \\
    $\bm{v}_{\rm k}$  & SN natal kick vector \\
    $v_{\rm k}$  & SN natal kick magnitude \\
    $\theta_{\rm k}$    & SN natal kick polar angle \\
    $\phi_{\rm k}$  & SN natal kick azimuthal angle \\
    $v_{\rm sys}$   & Systemic velocity \\
    $v_{\rm orb}$ & Orbital velocity \\
    $v_1$   & Post-kick primary velocity \\
    $\alpha$    & Right ascension \\
    $\delta$    & Declination \\
    $t$     & Birth time (system age) \\
    $x_{\rm type}$  & Type of system (e.g., HMXB) \\
    $\bm{D}$ & Set of observed properties \\
    $\bm{\mathcal{O}}$   &   Set of binary observables with uncertainties \\
    $\alpha_{\rm IMF}$    & IMF power law index \\
    $C_{\rm m}$ & IMF normalization constant \\
    $C_{\rm a}$ & Orbital separation normalization constant \\
    $Z$ & Metallicity \\
    $N_{\rm LMC}$   & Number of stars formed in the LMC \\
    $m_f$ & mass function \\
    $i$ & inclination angle \\
    $\omega$ & Polar angle for the projected velocity \\
    $d$ & Distance traveled by an HMXB \\
    $D_{\rm LMC}$   & Distance to the LMC \\
    $s$ & \vspace{-0.2cm} Projected separation between an HMXB\\
        & and its birth location \\
    $\theta_{\rm proj}$ & \vspace{-0.2cm} Polar angle between an HMXB \\
        & and its birth location \\
    $\theta_{\rm c}$    & \vspace{-0.2cm} Maximum angular separation \\
        & between an HMXB and its birth location \\
    $\phi$  & \vspace{-0.2cm} Azimuthal angle between an HMXB \\
        & and its birth location \\
    $J_{\rm coor}$  & \vspace{-0.2cm} Jacobian transformation matrix \\
        & between $(\alpha,\delta)$ and $(\theta,\phi)$ \\
    $L_{\rm x}$ & X-ray luminosity \\
    $\alpha_{\rm SN}$    & First kick constant \\
    $\beta_{\rm SN}$ & Second kick constant \\
\enddata
\end{deluxetable}

\subsection{Modeling Populations}
\label{sec:stats_population}

In BPS, one randomly produces a set of binaries by evolving the distribution of initial binary parameters through binary evolution prescriptions. Since in general we do not {\it a priori} know which initial conditions will produce systems of a certain population (for instance, many binaries are disrupted or merge during their evolution), we must test the entire region of initial binary parameter space that could plausibly produce that population. Traditional BPS codes solve this problem by making random draws of $\bm{x}_{\rm i}$, the initial binary parameters, from observationally derived distribution functions, $P(\bm{x}_{\rm i})$:
\begin{equation}
\bm{x}_{\rm i} \sim P(\bm{x}_{\rm i}).
\end{equation}
To first order, at a given metallicity, high-mass binaries can be determined uniquely by only a few parameters (ignoring dynamical effects such as three-body interactions): the binary components' initial masses, $M_{\rm 1,i}$ and $M_{\rm 2,i}$, the separation, $a_{\rm i}$, the eccentricity, $e_{\rm i}$, and the birth time $t_{\rm i}$. Depending on the population being modeled, we may optionally include the coordinates for the binary's birth position, $\alpha_{\rm i}$ and $\delta_{\rm i}$, and the kick velocity received when the primary, and potentially the secondary, collapsed to form a compact object:
\begin{equation}
\bm{x}_{\rm i} \equiv ( M_{\rm 1,i}, M_{\rm 2,i}, a_{\rm i}, e_{\rm i}, \bm{v}_{\rm k, SN1}, \bm{v}_{\rm k, SN2}, \alpha_{\rm i}, \delta_{\rm i}, t_{\rm i} ), \label{eq:x_i}
\end{equation}
where the two SN kicks are discerned by their separate subscripts. One key aspect of this model is that the SN kick magnitudes and directions are included as model parameters rather than determined from random draws on-the-fly during binary evolution.

Using a binary evolution code, these initial binaries are then evolved from $\bm{x}_{\rm i}$ into its current state, represented by $\bm{x}_{\rm f}$:
\begin{equation}
\bm{x}_{\rm f} = f(\bm{x}_{\rm i}). \label{eq:xf_xi}
\end{equation}
We can now define a function $P(x_{\rm type} \given \bm{x}_{\rm i})$ to be the indicator function, which is either unity or zero depending on whether $\bm{x}_{\rm f}$ represents a system of the specific type we are trying to model:
\begin{equation}
P(x_{\rm type} | \bm{x}_{\rm i}) = 
\begin{cases}
1, & \bm{x}_{\rm f} \in x_{\rm type} \\
0, & \bm{x}_{\rm f} \notin x_{\rm type},
\end{cases}
\label{eq:x_type}
\end{equation}
where $M$ is our binary evolution model. Distributions of the components of $\bm{x}_{\rm f}$ (such as the spatial distribution of systems, X-ray luminosity function of HMXBs, orbital period distribution, etc.) can then provide model predictions for populations or comparisons to observational samples.

For binary populations involving NSs and BHs, traditional BPS may be an inefficient tool; mass transfer and SN kicks may merge or disrupt the majority of systems before they evolve into objects of interest (i.e., $P(x_{\rm type} \given \bm{x}_{\rm i}) = 0$ for many or even most of the randomly drawn $\bm{x}_{\rm i}$). The fact that these binaries are discarded is a major source of computational expense in traditional BPS.

Rather than taking random draws of $\bm{x}_{\rm i}$, we consider the components of $\bm{x}_{\rm i}$ model parameters. In this formulation $P(\bm{x}_{\rm i})$ is the prior probability on the model parameters, and $P(x_{\rm type} \given \bm{x}_{\rm i})$ is the likelihood of producing a binary of a particular type from a given $\bm{x}_{\rm i}$. Using Bayes' Theorem, we can then identify the set of $\bm{x}_{\rm i}$ most likely to produce these binaries:
\begin{equation}
P(\bm{x}_{\rm i} \given x_{\rm type}) = \frac{P(x_{\rm type} \given \bm{x}_{\rm i}) P(\bm{x}_{\rm i})} {P(x_{\rm type})}. \label{eq:bayes_pop}
\end{equation}
We ignore $P(x_{\rm type})$, which serves as a normalization constant, and define the posterior probability as the numerator on the right hand side of Equation \ref{eq:bayes_pop}.\footnote{Although in this work we ignore it, the denominator is required to determine absolute formation rates for evolutionary channels. We discuss this further in Section \ref{sec:limitations} and Appendix \ref{sec:evidence}.} The large dimensionality of $\bm{x}_{\rm i}$ argues for an efficient numerical method to probe the region of viable parameter space.

In an MCMC algorithm, a ``walker'' moves around the $\bm{x}_{\rm i}$ parameter space: the posterior probability of the current $\bm{x}_{\rm i}$ is calculated, a new trial $\bm{x}_{\rm i}$ is randomly selected, the posterior probability of the new position is compared to that of the current position, and depending on the ratio of the two posterior probabilities, the new $\bm{x}_{\rm i}$ is either selected and added to the chain or rejected and the current position is kept for another step. The chain stores a record of all the walker's past positions. Samples from this chain comprise the synthetic population analogous to the population generated by traditional BPS.\footnote{Since the current walker position will necessarily be closely related to the previous step, the posterior samples will be correlated with some characteristic length. The autocorrelation length needs to be calculated {\it a posteriori}, and only one sample per autocorrelation length can be considered independent. We discuss the calculation of autocorrelation lengths in Appendix \ref{sec:initialize}.}

In principle, for an infinite number of iterations, the distribution of posterior samples of $\bm{x}_{\rm i}$ produced by this method will identically mimic the distribution generated by traditional BPS. Since we are limited to a finite sample, the computation time to produce a statistically robust sample using each method depends on the relative formation efficiency of systems and the autocorrelation length of the MCMC posterior distribution. For systems with a high formation efficiency, traditional methods may be preferred, since every random draw from BPS is independent. However for systems with a low formation efficiency or short lifetime, MCMC may be preferred.

Correct implementation requires careful attention to the prior distributions, $P(\bm{x}_{\rm i})$, and an efficient method to calculate $P(x_{\rm type} \given \bm{x}_{\rm i})$. We describe how we calculate the prior probabilities in Section \ref{sec:priors} and our binary evolution prescription, which is a modified version of \bse, in Appendix \ref{sec:pybse}.

\subsection{Modeling Individual HMXBs} \label{sec:stats_individual}

If we would like to quantitatively compare a model to a set of observationally derived properties, $\bm{D}$, of a particular system rather than a population, we need to adapt our method. We may be interested in either deriving the initial binary conditions that could have produced the observed systems, $P(\bm{x}_{\rm i} \given \bm{D})$, or determining the current binary parameters, $P(\bm{x}_{\rm f} \given \bm{D})$. These two quantities are closely related since binary evolution directly relates $\bm{x}_{\rm i}$ to $\bm{x}_{\rm f}$.

Instead of calculating $P(\bm{x}_{\rm i} \given x_{\rm type})$ as in Equation \ref{eq:bayes_pop}, we can again use Bayes' Theorem to calculate $P(\bm{x}_{\rm i} \given \bm{D})$:
\begin{equation}
P(\bm{x}_{\rm i} \given \bm{D}) = \frac{P(\bm{D} \given \bm{x}_{\rm i})P(\bm{x}_{\rm i})}{P(\bm{D})}, \label{eq:bayes_ind}
\end{equation}
where $P(\bm{D} \given \bm{x}_{\rm i})$ is the likelihood function, $P(\bm{x}_{\rm i})$ is the prior probability on the model parameters, and $P(\bm{D})$ is again a normalization constant that for our purposes can be ignored. The posterior probability is the numerator in the right hand side of Equation \ref{eq:bayes_ind}.

Traditional BPS takes a shotgun approach, making many random draws of $\bm{x}_{\rm i}$ from $P(\bm{x}_{\rm i})$. If one wants to then calculate a Bayesian posterior probability, $P(\bm{x}_{\rm i} \given \bm{D})$ can be determined from the subset of systems that are consistent with the observations \citep[e.g.,][]{andrews15}. Results are then derived from the selected subset of systems. Since only a small subset of the binaries that form a type of binary of interest will be consistent with any particular observed system, the likelihood function may be non-zero for only a small region of parameter space. The more precisely a system is measured, the smaller the phase space volume of interest.

We simulate individual systems using the same model parameters that we use for a population of systems, so $\bm{x}_{\rm i}$ is still defined by Equation \ref{eq:x_i}. The priors on these parameters are identical for individual systems compared with those derived for populations of HMXBs.

Individual systems may have well measured quantities such as the orbital period ($P'_{\rm orb}$), eccentricity ($e'$) or companion mass ($M'_2$), where primed quantities indicate observed rather than true, underlying values. Each of these measured quantities has some uncertainty associated with it which should be taken into account. For ease of notation, we will combine the set of these observables into $\bm{\mathcal{O}}'$. Furthermore, individual systems have a specific observed location that we are trying to associate with nearby star forming regions. In the following discussion, we will assume that we are including the system's current position as an observable, but this constraint may be trivially removed.

We start by defining $\bm{D}$ as:
\begin{equation}
\bm{D} \equiv (\alpha, \delta, \bm{\mathcal{O}}', x_{\rm type}). \label{eq:D}
\end{equation}
Uncertainties on the observed quantities are not explicitly included in $\bm{D}$, and we ignore uncertainties on the position. To generate our likelihood function, we now marginalize over the true values of the observables ($\bm{\mathcal{O}}$) and the (scalar) systemic velocity ($v_{\rm sys}$). Our model likelihood then becomes:
\begin{equation}
P(\bm{D} \given \bm{x}_{\rm i}) = \int \dd \bm{\mathcal{O}}\ \dd v_{\rm sys}\ P( \bm{\mathcal{O}}, v_{\rm sys}, \bm{D} \given \bm{x}_{\rm i}).
\end{equation}
We substitute for $\bm{D}$, and based on independence we factor the integrand into separate, tractable parts; the observed quantities, $\bm{\mathcal{O}}'$, are dependent only on their underlying values, $\bm{\mathcal{O}}$, the current binary parameters are dependent on the initial conditions and the binary evolution model, and the binary's observed position is dependent on the systems' initial conditions (which contains its age and birth location) and its velocity:
\begin{eqnarray}
P(\bm{D} \given \bm{x}_{\rm i}) &=& \int \dd \bm{\mathcal{O}}\ \dd v_{\rm sys}\ P(\bm{\mathcal{O}}' \given \bm{\mathcal{O}})\ P(\bm{\mathcal{O}}, v_{\rm sys} \given \bm{x}_{\rm i}) \nonumber \\
  & & \qquad \times P(\alpha, \delta \given \bm{x}_{\rm i}, v_{\rm sys})\ P(x_{\rm type} \given \bm{x}_{\rm i}). \label{eq:marginalized}
\end{eqnarray}

The first term in the integrand of Equation \ref{eq:marginalized}, $P(\bm{\mathcal{O}}' \given \bm{\mathcal{O}})$, accounts for the observational uncertainties on the binary's parameters.
We discuss this term along with the second term in the integrand, which describes the function evolving the binary from its {\it ab initio} state to the parameters of the system today, in Section \ref{sec:likelihood_observables}. It is worth noting that this is the only term which depends on the chosen set of binary evolution prescriptions.

The position term in the integrand accounts for the fact that the system's birth place will, in general, be different from its observed position since the center of mass of a system received a kick during the primary's core collapse. We explicitly include the dependence on $\bm{x_{\rm i}}$ and $v_{\rm sys}$ since the distance travelled depends on both the system's velocity and the time since the primary's SN. We derive this term in Section \ref{sec:ra_dec} below.

The final term in the integrand in Equation \ref{eq:marginalized} is the same indicator function provided in Equation \ref{eq:x_type} that is used to model populations of systems. This function ensures that the likelihood is non-zero only for $\bm{x}_{\rm i}$ that produce systems of the same abstract class (HMXBs throughout this work) as our observed binary.

\subsection{Prior Probabilities: $P(\bm{x}_{\rm i})$} 
\label{sec:priors}

Typically, prior probabilities are set within rapid BPS codes. In \dart, a BPS code is used only to rapidly evolve binaries from their initial conditions. User-defined, external prior probabilities for individual parameters can be easily imported. Here, we describe the default prior probabilities, which are used in the tests and examples in the remainder of this work. Our model includes between eight and 13 parameters, which can be factored into several parts:
\begin{eqnarray}
P(\bm{x}_{\rm i}) &=& P(M_{\rm 1,i})\ P(M_{\rm 2,i}\given M_{\rm 1,i})\ P(a_{\rm i} \given e_i)\ P(e_{\rm i}) \nonumber \\
  & & \qquad \times P(\bm{v}_{\rm k,1})\ P(\bm{v}_{\rm k,2})\ P(\alpha_{\rm i}, \delta_{\rm i}, t_{\rm i})
\end{eqnarray}
This equation is equivalent for the $P(\bm{x}_{\rm i})$ term used in both Equations \ref{eq:bayes_pop} and \ref{eq:bayes_ind}. We discuss the priors on each model parameter in turn below.

\subsubsection{Initial Binary Parameters}
\label{sec:priors_binary}

Although more modern, observationally calibrated distributions exist \citep[e.g.,][]{moe17}, our prior probabilities over $M_{\rm 1,i}$, $M_{\rm 2,i}$, $a_{\rm i}$, and $e_{\rm i}$ are all equivalent to or similar to the distributions used in most previous population synthesis studies of high mass binaries \citep[e.g.,][]{belczynski08}.

The initial primary mass follows a power law initial mass function (IMF):
\begin{equation}
P(M_{\rm 1,i}) = C_{\rm m} M_{\rm 1,i}^{\alpha_{\rm IMF}};\ M_{\rm 1,i} \in [M_{\rm 1,min}, M_{\rm 1,max}]
\end{equation}
where $C_{\rm m}$ is a normalization constant dependent upon the limits of the distribution ($M_{\rm 1,min} = 8 \Msun$ and $M_{\rm 1,max} = 150 \Msun$) and $\alpha_{\rm IMF}$:
\begin{equation}
C_{\rm m} = \frac{\alpha_{\rm IMF} + 1}{M_{\rm 1,max}^{\alpha_{\rm IMF}+1} - M_{\rm 1,min}^{\alpha_{\rm IMF}+1}}.
\end{equation}
We choose 8 \Msun\ as the lower mass limit typically producing NSs, and 150 \Msun\ as the upper mass limit. In the present analysis, since the distribution strongly preferences lower mass stars, our qualitative results are independent of the upper mass limit. We choose a Salpeter power law: $\alpha_{\rm IMF} = -2.35$ \citep{salpeter55}.

We choose a prior on the secondary mass based on a flat mass-ratio distribution which has the subtle effect that the prior on the secondary is dependent on that of the primary. The maximum mass-ratio is unity to ensure the primary is the more massive of the pair, and the minimum secondary mass is set to 2 \Msun. This leads to a prior probability:
\begin{equation}
P(M_{\rm 2,i} \given M_{\rm 1,i}) = \frac{1}{M_{\rm 1,i} - 2\ \Msun};\ M_{\rm 2,i} \in [2.0\ \Msun, M_{\rm 1,i}]
\end{equation}

We choose a thermal initial eccentricity distribution \citep{duquennoy91}:
\begin{equation}
P(e_{\rm i}) = 2e_{\rm i};\ e_{\rm i} \in[0,1].
\end{equation}

Finally, we choose a prior on the initial orbital separation of the binary that scales with $a_{\rm i}^{-1}$ \citep{abt83}:
\begin{equation}
P(a_i \given e_i) = \frac{C_{\rm a}}{a_{\rm i}};\ a_{\rm i} \in [a_{\rm min}, a_{\rm max}],
\end{equation}
where $C_{\rm a}$ is a normalization constant
\begin{equation}
C_{\rm a} = \frac{1}{\log a_{\rm max} - \log a_{\rm min}},
\end{equation}
and $a_{\rm min}$ and $a_{\rm max}$ are set so that the system will not be separated by less than 10 $R_{\odot}$ at pericenter or more than 10$^4 R_{\odot}$ at apocenter (hence the dependence on $e_i$).

\subsubsection{SN Kick Parameters}

The SN kick velocity, $\bm{v}_{\rm k}$, is composed of three parameters, which we represent as a kick magnitude ($v_{\rm k}$) and two angles ($\theta_{\rm k}, \phi_{\rm k}$)\footnote{Typically, tides and mass transfer have circularized the binary prior to core collapse. In the rare cases in which a binary has not yet circularized, a fourth parameter is needed: the mean anomaly of the orbit corresponding to the orbital phase at which core collapse occurs. For simplicity, we currently ignore this complication.}. If the binary is a double compact object, the two SNe are independent, but have the same prior probabilities. In traditional BPS, SN kick parameters are determined through Monte Carlo random draws from a predefined distribution on-the-fly during each binary's evolution. In our model, we instead include the SN kick magnitude and direction as model parameters with prior probabilities corresponding to standard distributions: we assume that $v_{\rm k}$ follows a Maxwellian distribution with a dispersion of 265 km s$^{-1}$ \citep{hobbs05}. We can therefore express the normalized probability of $v_{\rm k}$ as:
\begin{equation}
P(v_{\rm k}) = \sqrt{\frac{2}{\pi}} \frac{v_{\rm k}^2} {\sigma^3} {\rm exp} \left[ -v_{\rm k}^2 / 2 \sigma^2 \right];\ v_{\rm k} \in [0, \infty). \label{eq:P_v_k}
\end{equation}

Since the kick distribution is assumed to be isotropic, normalized probabilities for the kick polar, $\theta_{\rm k}$, and azimuthal, $\phi_{\rm k}$, angles are straightforward:
\begin{eqnarray}
P(\theta_{\rm k}) &=& \frac{\sin \theta_{\rm k}}{2};\ \theta_{\rm k} \in [0, \pi] \label{eq:P_theta_k} \\
P(\phi_{\rm k}) &=& \frac{1}{\pi};\ \phi_{\rm k} \in [0, \pi] . \label{eq:P_phi_k}
\end{eqnarray}
Formally, $\phi_{\rm k}$ varies between 0 and $2 \pi$; however, as described in Appendix \ref{sec:supernova}, the only contribution of $\phi_{\rm k}$ to the evolution of the binary is through a $\sin^2 \phi_{\rm k}$ term, which is periodic from 0 to $\pi$.

\subsubsection{Star Formation History}

\begin{figure*}
\begin{center}
\includegraphics[width=0.95\textwidth]{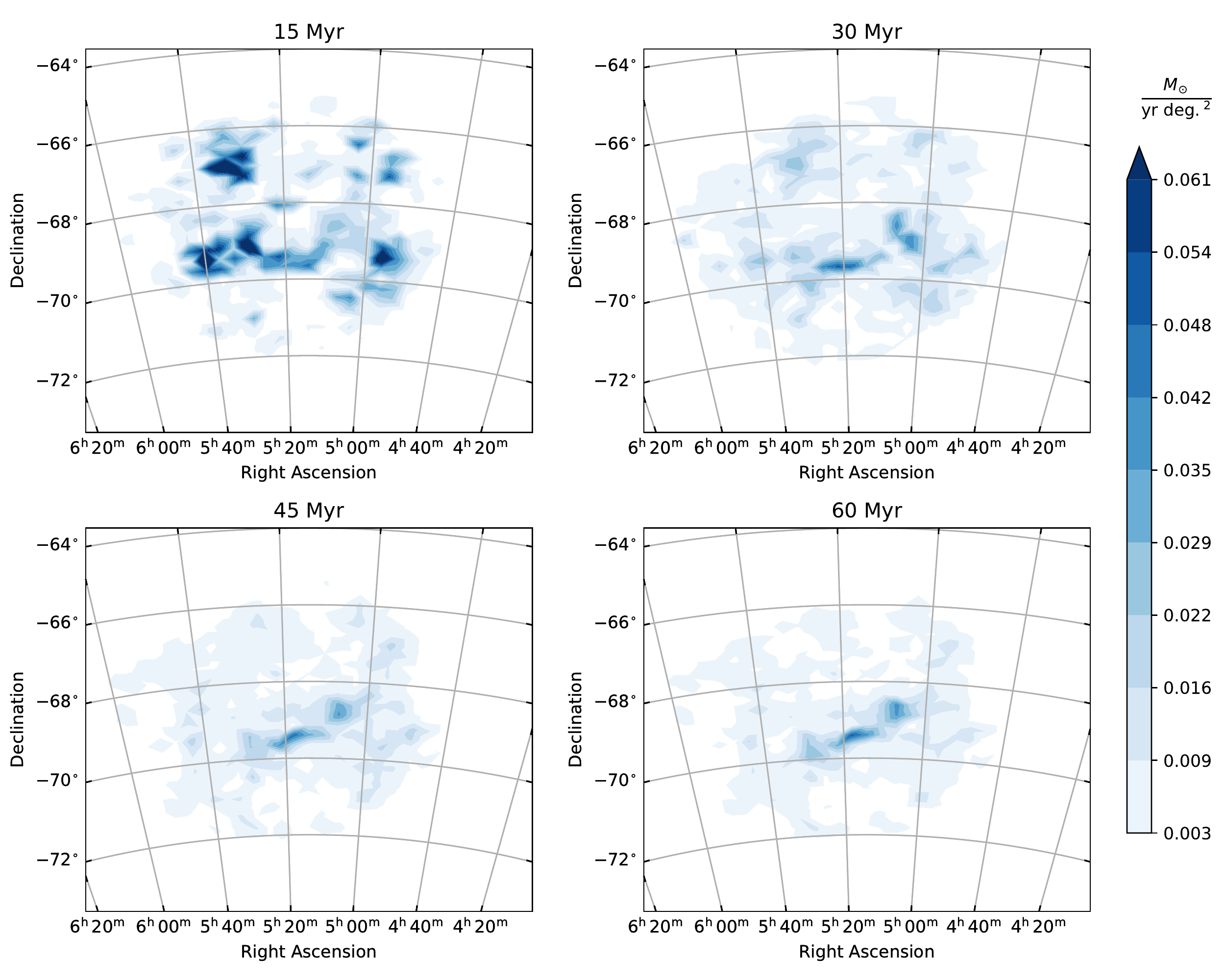}
\caption{The prior on both position of the binary's birth location and time depends on the LMC's star formation history maps produced by \citet{harris09}. We show samples of the SFH at four different ages spanning the range of typical HMXB lifetimes. These demonstrate the typical resolution of the spatially resolved SFH. Note that the LMC experienced a period of rapid star formation in the past $\approx$20 Myr.}
\label{fig:LMC_SFH}
\end{center}
\end{figure*}

The priors on $\alpha_{\rm i}$, $\delta_{\rm i}$, and $t_{\rm i}$ depend on the local SFH at that position and time. Our model can be run with a basic time-dependent SFH, ignoring the systems' positions, but the most power is gained by including spatially resolved SFH maps as a prior on both position and time. One example is the SFH map for the LMC from \citet{harris09}. These maps cover the LMC with $\sim$1300 separate regions with angular resolutions of 12\amin\ on a side in the inner regions and 24\amin\ on a side in the outer regions. The SFH for each region has a resolution of 0.2 dex in $t$ ranging from 6.8 to 10.2 in log $t$. We ignore uncertainties on the SFHs and generate linear interpolation functions over log $t$ for each of the 1300 regions. For testing with HMXBs, we only take into account star formation at a metallicity of $Z=0.008$, the dominant metallicity at which stars have been formed in the LMC over the past 1 Gyr. Since HMXBs have all been born in the past 10$^8$ yrs, the older, lower-$Z$ population is irrelevant for the systems we model in this work.

Of course, not every LMC star formed in the past 100 Myr had a metallicity exactly equal to 0.008, however the fits from \citet{harris09} indicate that their choice of metallicities provides a good approximation to the observations. A more complete model will include $Z$ as a model parameter, accounting for the (spatially dependent) metallicity evolution of the LMC. We leave this for future work, although we note that deviations in the metallicity of order 0.1-0.2 dex are not expected to significantly alter the HMXB population \citep{fragos13a, fragos13b}.

These spatially resolved SFHs provide the function: ${\rm SFR}(\alpha_{\rm i}, \delta_{\rm i}, t_{\rm i})$, the rate per unit area on the sky that stars were formed at a specific location and time in the LMC. With a normalization constant, this spatially dependent star formation rate is the prior on position and time:
\begin{equation}
P(\alpha_{\rm i}, \delta_{\rm i}, t_{\rm i}) = \frac{1}{M_{\rm LMC}} {\rm SFR}(\alpha_{\rm i}, \delta_{\rm i}, t_{\rm i}),
\end{equation}
where $M_{\rm LMC}$ is the combined mass of stars with $Z=0.008$ produced throughout the lifetime of the LMC. Figure \ref{fig:LMC_SFH} shows the star formation rate map at four different ages spanning the range of typical HMXB lifetimes. These maps indicate that the locations and overall rate of star formation have substantially evolved over the past 60 Myr.

\subsection{Binary Parameter Likelihood} \label{sec:likelihoods}

Given a birth time and a particular set of initial binary parameters, the likelihood function provides the probability that a binary of interest will be formed. For populations of systems, the likelihood is simply the function provided in Equation \ref{eq:x_type}. Determining this function nevertheless requires evolving the system through its evolution. Currently, this requires rapid binary evolution codes. Although \dart is developed to be used with any binary population synthesis code, we provide a modified version of one such code, \bse, along with \dart. Our minor modifications to the freely available version are described in Appendix \ref{sec:pybse}.

\subsubsection{Fitting Observables with Uncertainties}
\label{sec:likelihood_observables}

If we would like to model an individual system with a set of observations, the likelihood function includes the observations and their uncertainties. For evolved stellar binaries, calculating this term relies first on the evolution from $\bm{x}_{\rm i}$ to $\bm{x}_{\rm f}$ which is determined by a rapid binary evolution code. Using our example of HMXBs, the rapid binary evolution code provides the term $P(\bm{\mathcal{O}}, v_{\rm sys} \given \bm{x}_{\rm i})$ in Equation \ref{eq:marginalized}. We then need to compare the model results with the observed properties of the system. For an observed parameter with a known, Gaussian standard deviation (measurement uncertainty), the likelihood function is straightforward and involves the evaluation of the probability density of the normalized Gaussian distribution for that parameter at its final state, $\mathcal{O}$:
\begin{equation}
P(\mathcal{O}'|\mathcal{O}) = \mathcal{N}(\mathcal{O}' | \mathcal{O}, \sigma_{\mathcal{O}}),
\end{equation}
where $\mathcal{N}$ represents a normalized Gaussian distribution with mean $\mathcal{O}$ and standard deviation $\sigma_{\mathcal{O}}$, evaluated at $\mathcal{O}'$: 
\begin{equation}
\mathcal{N}(\mathcal{O}' | \mathcal{O}, \sigma_{\mathcal{O}}) =  \frac{1}{\sqrt{2 \pi \sigma_{\mathcal{O}}^2}} \exp\left[ - \frac{(\mathcal{O} - \mathcal{O}')^2} {2 \sigma^2_{\mathcal{O}}} \right]. 
\end{equation}

Our method can adapt to any observation of an individual stellar parameter included in our model; we only need to compare the evolved binary parameters to the observations. To simultaneously fit multiple observations, the resulting likelihood is then a product of the fits to each of the $k$ individual observed quantities:
\begin{equation}
P(\bm{\mathcal{O}}' | \bm{\mathcal{O}}) = \prod_k \mathcal{N}(\mathcal{O}'_k | \mathcal{O}_k, \sigma_{\mathcal{O}_k}).
\end{equation}

Observables may not always be in a Gaussian form. For instance, in some cases only an upper limit is measured for a binary's eccentricity: $e'_{\rm upper}$. In this case, the likelihood function for the eccentricity term is:
\begin{equation}
P(e'_{\rm upper} | e) = 
\begin{cases}
e_{\rm upper}^{-1}, & e < e'_{\rm upper} \\
0, & e > e'_{\rm upper},
\end{cases}
\label{eq:e_upper}
\end{equation}
The $e_{\rm upper}^{-1}$ term is a normalization constant.

Other observables may have half-bounded constraints. Examples may include upper or lower limits on metallicity measurements, [Fe/H], or on the log of the X-ray luminosity, log L$_{\rm x}$. Formally, in such cases the support for $P(\mathcal{O}_{\rm upper/lower}' \given \mathcal{O})$ may extend infinitely in one direction, and the likelihood function over $\mathcal{O}$ cannot be normalized. This problem can be avoided by transforming these measurement constraints from a limit on $\mathcal{O}$ to a limit on $\exp[\pm \mathcal{O}]$, however one must make careful choices about whether the likelihood function over $\exp[\pm \mathcal{O}]$ should also be uniform within the allowed range. Because the model parameter space is explored by our MCMC algorithm based only on the ratio of posterior probabilities, the results are not affected by constant coefficients in the likelihood. In cases with half-bounded constraints, we therefore recommend using the Heaviside step function:
\begin{equation}
P(\mathcal{O}_{\rm upper}' \given \mathcal{O}) = H[\mathcal{O}_{\rm upper}' - \mathcal{O}] \equiv \begin{cases}
1, & \mathcal{O} < \mathcal{O}_{\rm upper}' \\
0, & \mathcal{O} > \mathcal{O}_{\rm upper}'.
\end{cases}
\end{equation}
An analogous constraint can be made for corresponding lower limits.

\subsubsection{Mass Function}
\label{sec:mass_function}

In the case of many binary systems, one is limited to observations of the mass function, $m_f$, rather than the companion mass. $m_f$ depends on the inclination angle, $i$, by which the system is viewed, and comparing simulated systems to an observed mass function requires a convolution integral over the inclination angle. Since the measured $m_f$ (denoted as $m'_f$) typically has some uncertainty associated with its measurement, we further require an additional convolution integral over the {\it true} mass function $m_f$: 
\begin{equation}
P(m'_f \given \bm{x}_{\rm i}) = \int \dd m_f \int \dd i\ P(m_f, m'_f, i \given \bm{x}_{\rm i}).
\end{equation}

After separating terms, we have:
\begin{equation}
P(m'_f \given \bm{x}_{\rm i}) = \int \dd m_f \int \dd i\ P(m'_f \given m_f)\ P(m_f \given i, \bm{x}_{\rm i})\ P(i) \label{eq:m_f_tmp}
\end{equation}
where the distribution over inclination angles is $P(i)\sim \sin i$, $P(m'_f \given m_f)$ is the measurement and uncertainty associated with the mass function (typically modeled as a Gaussian), and $P(m_f \given i, \bm{x}_{\rm i})$ is a delta function derived from the definition of $m_f$. The delta function reduces the double integral in Equation \ref{eq:m_f_tmp} to a single integral:
\begin{equation}
P(m'_f \given \bm{x}_{\rm i}) = \int \dd m_f\ P(m'_f \given m_f)\ h(M_1, M_2, m_f), \label{eq:mass_function}
\end{equation}
where, from \citet{andrews14}:
\begin{equation}
h(M_1, M_2, m_f) = \frac{(M_1+M_2)^{4/3}}{3\ m_f^{1/3}M_2\sqrt{M_2^2 - \left[ m_f(M_1+M_2)^2 \right]^{2/3}}}. 
\end{equation}

\dart\ includes the mass function (and its associated uncertainty) as an optional observable. The integral in Equation \ref{eq:mass_function} is calculated using a Monte Carlo method:
\begin{equation}
P(m'_f \given \bm{x}_{\rm i}) \approx \frac{1}{N} \sum_{k=1}^N h(M_1, M_2, m_{f,k}) 
\end{equation}
where we draw $N$ random samples of $m_{f,k}$ from the Gaussian distribution centered around the observed $m_f$ with its associated uncertainty, $\sigma_{m_f}$:
\begin{equation}
m_{f,k} \sim \mathcal{N}(m_f \given m'_f, \sigma_{m_f}).
\end{equation}
We find that 100,000 random draws provide a sufficiently precise calculation of this integral.

\subsubsection{Position Likelihood} 
\label{sec:ra_dec}

The position likelihood provides the probability that, given a system's position, systemic velocity, and time since SN, the system would be observed at its current position. 
To solve the positional component of Equation \ref{eq:marginalized}, we first marginalize over $\omega$, the angle between the line of sight vector to the birth location and the systemic velocity vector:
\begin{equation}
P(\alpha, \delta \given \bm{x}_{\rm i}, v_{\rm sys} ) = \int \dd \omega\ P(\alpha, \delta, \omega \given \bm{x}_{\rm i}, v_{\rm sys} ). \label{eq:P_pos_1}
\end{equation}

\begin{figure}
\begin{center}
\includegraphics[width=0.95\columnwidth]{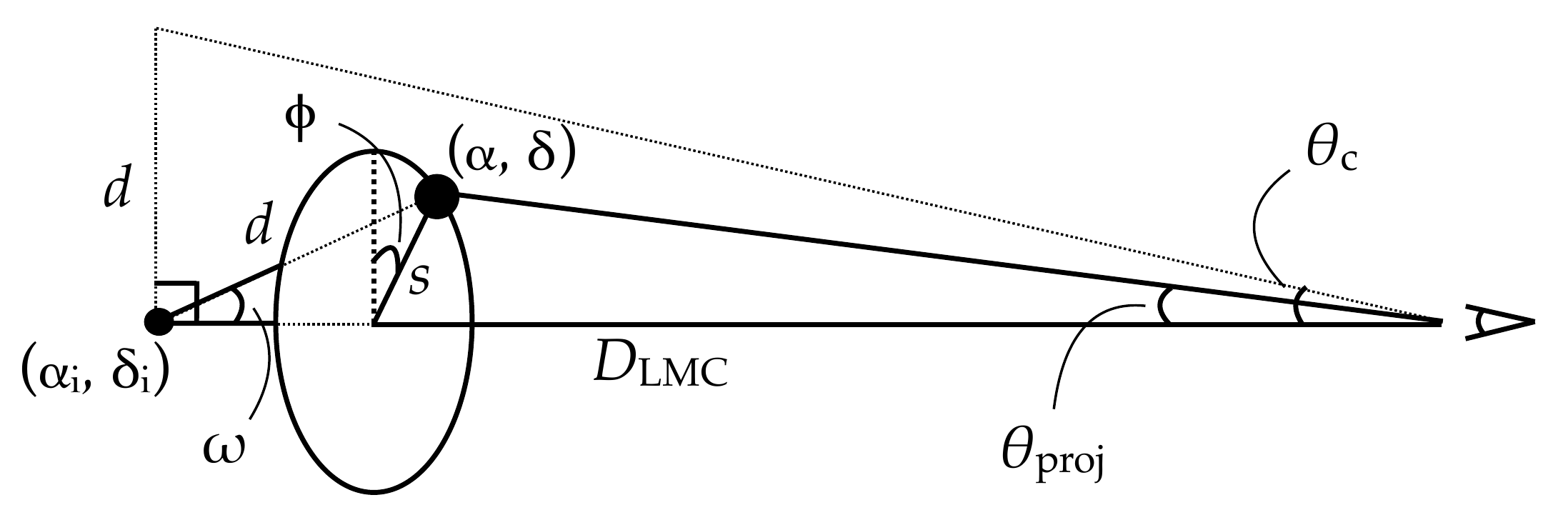}
\caption{Our representation of the current position on the plane of the sky $(\alpha, \delta)$ in relation to its birth position $(\alpha_{\rm i}, \delta_{\rm i})$. The distance the system traveled is $d$, which has a projected separation $s$. We express this transformation as a function of angular separation, $\theta_{\rm proj}$, and position angle, $\phi$. Note, for typical systems $d << D_{\rm LMC}$.}
\label{fig:position_projection}
\end{center}
\end{figure}

We next perform a coordinate transformation from the absolute positional coordinates $\alpha$ and $\delta$ to the relative angular separation, $\theta_{\rm proj}$, and the position angle, $\phi$, measured from the system's birth location. Figure \ref{fig:position_projection} shows our parameterization of the transformation from a system's birth location at $\alpha_{\rm i}$ and $\delta_{\rm i}$ to its current location at $\alpha$ and $\delta$. Standard formulae defining the angular separation and position angle of double stars can be used:
\begin{eqnarray}
\theta_{\rm proj} &\approx& \sqrt{(\alpha_{\rm i}-\alpha)^2 \cos\delta_{\rm i} \cos\delta + (\delta_{\rm i}-\delta)^2} \\
\phi &\approx& \tan^{-1} \left[ \frac{\delta_{\rm i}-\delta}{(\alpha_{\rm i}-\alpha) \cos\delta_{\rm i}} \right].
\end{eqnarray}
The determinant of the Jacobian matrix for the transformation from ($\theta_{\rm proj}$, $\phi$) to ($\alpha$, $\delta$) can be calculated from partial derivatives:
\begin{eqnarray}
J_{\rm coor} &=& \left| \frac{\partial \theta_{\rm proj}}{\partial \alpha} \frac{\partial \phi}{\partial \delta} - \frac{\partial \theta_{\rm proj}}{\partial \delta}\frac{\partial \phi}{\partial \alpha}  \right| \nonumber \\
&\approx& \left| \frac{\cos \delta_{\rm i}}{\theta_{\rm proj}} \right|.
\end{eqnarray}

Equation \ref{eq:P_pos_1} now becomes:
\begin{eqnarray}
P(\alpha, \delta \given \bm{x}_{\rm i}, v_{\rm sys} ) &=& \int \dd \omega\ P(\theta_{\rm proj}, \phi, \omega \given \bm{x}_{\rm i}, v_{\rm sys} )\ J_{\rm coor} \nonumber \\
&=& \int \dd \omega\ P(\theta_{\rm proj} \given \omega,  \bm{x}_{\rm i}, v_{\rm sys} )\ P(\phi) \nonumber \\
& & \qquad \times P(\omega)\ J_{\rm coor}, \label{eq:P_pos_2}
\end{eqnarray}
where we have separated terms based on independence. $\omega$ is a randomly chosen polar angle and $\phi$ is a randomly chosen azimuthal angle: 
\begin{eqnarray}
P(\omega) &=& \frac{\sin \omega} {2};\ \omega \in [0,\pi] \\
P(\phi) &=& \frac{1}{2 \pi};\ \phi \in [0, 2\pi].
\end{eqnarray}

The physical distance a system travels is the product of $v_{\rm sys}$ and the time since the primary's core collapse, $t_{\rm travel}$:
\begin{equation}
d = v_{\rm sys} t_{\rm travel}.
\end{equation}
We ignore the effects of the host galaxy's gravitational potential and assume that systems move unperturbed in space after receiving a kick. We can only observe the projection of $d$ onto the sky, $s = d \sin \omega$. Separately, we can approximate $s$ as the product of $D_{\rm LMC}$ and $\theta_{\rm proj}$. After equating these two expressions for $s$ and solving for $\theta_{\rm proj}$, the first term of the integrand in Equation \ref{eq:P_pos_2} becomes a delta function:
\begin{equation}
P(\theta_{\rm proj} \given \omega, \bm{x}_{\rm i}, v_{\rm sys}) = \delta \left[G(\omega)\right], \label{eq:P_theta_proj}
\end{equation}
where:
\begin{equation}
G(\omega) = \theta_{\rm proj} - \frac{v_{\rm sys} t_{\rm travel} \sin \omega}{D_{\rm LMC}}.\end{equation}

With the delta function from Equation \ref{eq:P_theta_proj}, the integral in Equation \ref{eq:P_pos_2} can be reduced:
\begin{equation}
\int \dd \omega\ P(\phi) P(\omega) \delta \left[ G(\omega) \right]  J_{\rm coor}\  =\ \sum_j\ \frac{P(\omega_j^{\star}) P(\phi)  J_{\rm coor}}{ \left| \frac{ \dd G (\omega) }{\dd \omega} \right|_{\omega_j^*}},
\end{equation}
where the sum is over the roots of $G(\omega)$, $\omega_j^*$. There are two roots corresponding to whether the object is in front of or behind its birth location. This integral can now be evaluated analytically:
\begin{equation}
P(\alpha, \delta \given \bm{x}_{\rm i}, v_{\rm sys} ) =
\begin{cases} 
      0, & \theta_{\rm proj} \geq \theta_{\rm C}\\
     \frac{\tan \omega^*}{2 \pi \theta_{\rm C}}  J_{\rm coor}, & \theta_{\rm proj} < \theta_{\rm C} 
   \end{cases}
\end{equation}
where:
\begin{equation}
\omega^{\star} = \sin^{-1} \left[ \frac{\theta_{\rm proj}}{\theta_{\rm C}} \right].
\end{equation}

Note that we can alternatively express the non-zero part of $P(\alpha, \delta \given \bm{x}_{\rm i}, v_{\rm sys} )$ without the need for trigonometric functions:
\begin{equation}
    \frac{\tan \omega^*}{2 \pi \theta_{\rm C}}  J_{\rm coor} = \frac{1}{2 \pi \theta_{\rm C}} \frac{\theta_{\rm proj}}{\sqrt{\theta_{\rm C}^2 - \theta_{\rm proj}^2}} J_{\rm coor}.
\end{equation}

\subsubsection{X-ray Luminosity}
\label{sec:L_x}

For many HMXBs, the accretion-powered X-ray luminosity ($L_x$) is a principal observable. We calculate this luminosity using the following equation:
\begin{equation}
L_x = \eta \epsilon \frac{\mathcal{G}  M_{\rm acc} \dot{M}_{\rm acc}}{R_{\rm acc}},
\end{equation}
where $M_{\rm acc}$, $\dot{M}_{\rm acc}$, and $R_{\rm acc}$ are the accretor mass, accretion rate, and accretion radius, respectively, and are all provided by the rapid binary evolution code. $\epsilon$ is an accretion efficiency, and $\eta$ is the bolometric correction, accounting for the fraction of energy emitted in the observed X-ray band. \bse\ provides the accretion rate and accretor mass. For NS accretors, we use an accretion radius of 10 km, an accretion efficiency of unity, and a bolometric correction of 0.15, whereas for a BH accretor, we use an accretion radius of three Schwarzschild radii, an accretion efficiency of 0.5 and a bolometric correction of 0.8. These values are identical to those used by \citet{belczynski08} for wind-fed accretion.

\subsection{Markov Chain Monte Carlo Algorithm}
\label{sec:MCMC}

Ignoring the evidence, which is a constant for all $\bm{x}_{\rm i}$, the product of the prior probabilities and likelihood function form the posterior probability for a point in initial parameter space. Many algorithms exist for efficiently exploring a multidimensional parameter space, and since it is modular, \dart\ is constructed to allow for the easy inclusion of different codes. However, most algorithms are not well-suited for the formation of binaries with compact objects. This is because the indicator function within the likelihood produces infinitely steep boundaries separating the viable region in parameter space from the region that does not form systems of interest. The posterior space is therefore somewhat pathological; steep prior probabilities are cut off by the infinitely sharp likelihood distribution in a high dimensional space with boundaries unknown {\it a priori}.

\begin{deluxetable*}{lcccccc}
\tablecaption{Input parameters for our three mock binaries described in Section \ref{sec:mock} along with our model results for these values, denoted by the median value and 68\% confidence interval. We refrain from providing model constraints on the birth locations, as these are better understood pictorially in Figures \ref{fig:mock_2} and \ref{fig:mock_3}. \label{tab:mock_input} }
\tablehead{
\colhead{} &
\multicolumn{2}{c}{Mock System 1} & 
\multicolumn{2}{c}{Mock System 2} &
\multicolumn{2}{c}{Mock System 3} \\
\colhead{$\bm{x}_{\rm i}$} &
\colhead{Input} &
\colhead{Derived} &
\colhead{Input} &
\colhead{Derived} &
\colhead{Input} &
\colhead{Derived}
}
\startdata
$M_{\rm 1,i}$ ($M_{\odot}$) & 11.77 & 11.4$\substack{+1.8\\-0.7}$ & 14.11 & 12.4$\substack{+5.2\\-1.4}$ & 11.01 & 11.3$\substack{+0.8\\-0.5}$ \\
$M_{\rm 2,i}$ ($M_{\odot}$) & 8.07 & 7.6$\substack{+0.5\\-0.7}$  & 5.09 & 8.0$\substack{+2.0\\-2.4}$ & 7.42 & 7.5$\substack{+0.4\\-0.7}$ \\
$a_{\rm i}$ ($R_{\odot}$) & 4851 & 1280$\substack{+3650\\-1050}$  & 45 & 180$\substack{+2660\\-130}$ & 744 & 690$\substack{+1540\\-490}$ \\
$e_{\rm i}$ & 0.83 & 0.64$\substack{+0.26\\-0.27}$  & 0.62 & 0.57$\substack{+0.26\\-0.24}$ & 0.50 & 0.59$\substack{+0.26\\-0.25}$ \\
$v_{\rm k}$ (km s$^{-1}$) & 153 & 217$\substack{+157\\-93}$  & 141 & 211$\substack{+149\\-100}$ & 168 & 214$\substack{+190\\-115}$ \\
$\theta_{\rm k}$ (rad.) & 2.05 & 2.18$\substack{+0.48\\-0.51}$  & 1.70 & 2.16$\substack{+0.47\\-0.51}$ & 1.79 & 1.89$\substack{+0.57\\-0.52}$ \\
$\phi_{\rm k}$ (rad.) & 2.33 & 1.62$\substack{+0.90\\-0.97}$  & 1.63 & 1.58$\substack{+0.96\\-0.97}$ & 2.08 & 1.57$\substack{+0.53\\-0.55}$ \\
$\alpha_{\rm i}$ & - & - & 05:33:01.29 & - & 05:33:01.41 & - \\
$\delta_{\rm i}$ & - & - & -69:56:20.39 & - & -69:56:15.72 & - \\
$t_{\rm i}$ (Myr) & 34.74 & 35.9$\substack{+8.0\\-7.5}$ & 21.89 & 25.8$\substack{+8.0\\-8.1}$ & 36.59 & 30.0$\substack{+6.9\\-4.8}$ \\
\enddata
\end{deluxetable*}

Our preferred method to explore this parameter space based on the posterior probability in Equations \ref{eq:bayes_pop} or \ref{eq:bayes_ind} is the MCMC code {\tt emcee} \citep{foreman-mackey13}. {\tt emcee} employs an affine-invariant ensemble sampler using multiple ``walkers'' in concert \citep{goodman10}. We typically use 320 walkers. Since a large portion of the parameter space has a zero probability, it must be ensured that the walkers are initialized in non-zero probability regions of parameter space. Then the walkers must be ``burned-in'' until they have settled around the high probability region of parameter space. We provide details of our procedure for the initialization and burn-in of the walkers in Appendix \ref{sec:initialize}. To summarize, after initialization we run our simulation for a burn-in of at least 20,000 steps. We check the chains to make sure they have converged and run each model for at least 100,000 additional steps for sufficient statistics. The exact number of steps for our burn-in and production runs vary between models and should be determined separately on a case-by-case basis. Autocorrelation lengths and acceptance fractions also vary between the different models we test. We discuss these in detail in Appendix \ref{sec:initialize}.

In principle, the walkers could all be confined to a local maximum, while other maxima in the parameter space could exist. Such a situation would naturally arise if multiple evolutionary channels can form an individual system \citep[e.g., multiple evolutionary channels potentially forming Type Ia SNe;][]{claeys14} or if there are physically separate regions of star formation on the sky. In practice we find that {\tt emcee} is efficient at identifying multiple maxima, if they exist, even when they are separated by modest likelihood barriers. However, in some cases, particularly when modeling specific systems with precisely measured parameters, the likelihood barrier can be high enough that only the local maximum is explored that is closest to the initialization point. In our tests, we found this to be a problem only when applying our model to our third mock system which we discuss in Section \ref{sec:mock_3} and to the HMXB {\it Swift} J0513.4$-$6547 which we discuss in Section \ref{sec:J0513}. In these sections, we provide two different strategies for dealing with the problem of multiple maxima in the framework of \dart.

\section{Testing with Mock Systems}
\label{sec:mock}

Exactly because of the multidimensional nature of BPS, along with the potential for any particular system to be formed from different evolutionary channels, careful testing of our method is required. As a first test, we randomly choose three sets of initial conditions that produce HMXBs,
evolve these test systems forward using \bse, then attempt to recover the initial parameters using the current values of these parameters by adding uncertainties that mimic observational errors. Specifically, we test one system in which we ``observe" only the companion mass and eccentricity, one system in which we observe the sky position only, and one system with a precisely measured companion mass ($M_2$), orbital period ($P_{\rm orb}$), eccentricity ($e$), X-ray luminosity ($L_x$), and sky position, ($\alpha$, $\delta$). The exact values of the initial conditions and the observed quantities are provided in Tables \ref{tab:mock_input} and \ref{tab:mock_observed}, respectively. For the second and third mock systems, we use the LMC's SFH as a prior on the birth location and time.

\begin{figure*}
\begin{center}
\includegraphics[width=0.85\textwidth]{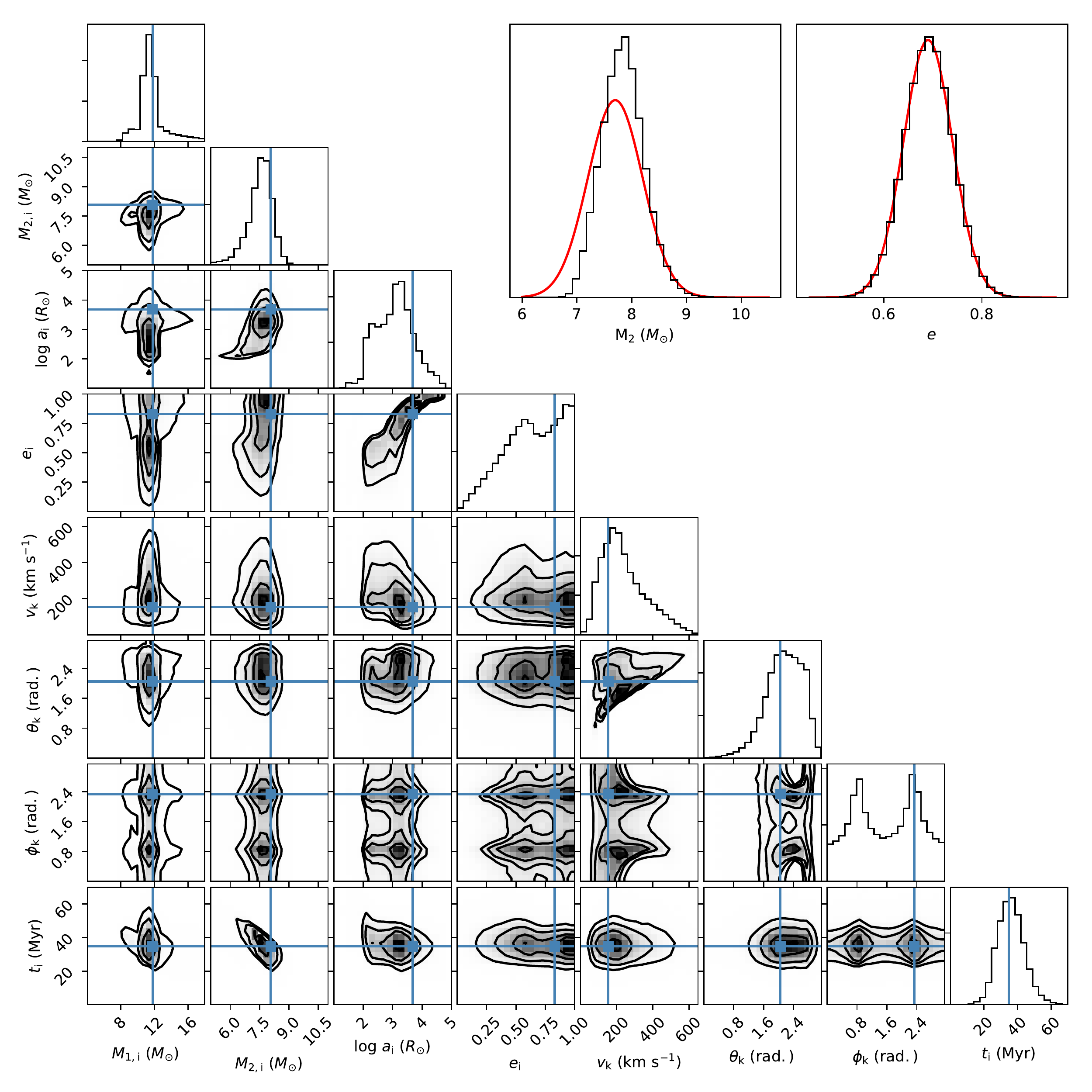}
\caption{ The covariances and 1D histograms of our posterior distribution of model parameters for mock system 1. The subscript i's on plot labels indicate initial binary parameters, each of which are described in Table \ref{tab:variable_index}. Our model successfully recovers all the input parameters. Top right panels compare the posterior distribution of current $M_2$ and $e$ (black distributions) with the ``observational'' constraints (red). These two distributions need not match exactly, but this comparison provides an important consistency check indicating that the model can, indeed, reproduce the observed parameters of the system.}
\label{fig:mock_1}
\end{center}
\end{figure*}

The principle product of applying our model to one of the mock systems, using the ``observed'' values provided in Table \ref{tab:mock_observed}, is a set of samples from the posterior distribution for the initial conditions of the binary. Comparison of these posterior samples to the input values for each of our mock systems provides a test of our algorithm. This is typically done by generating either 1D histograms of the distribution of individual parameters or contours for 2D projections of the higher dimensional parameter space of the different combinations of parameters. For example, Figure \ref{fig:mock_1} shows both the 1D histograms and the 2D distributions for all possible permutations of two of the eight initial binary parameters for mock system 1. Input values are designated by the blue horizontal and vertical lines in each panel.

\begin{deluxetable}{lccc}
\tablecaption{``Observed" quantities for our three mock binaries described in Section \ref{sec:mock}. \label{tab:mock_observed}}
\tablehead{
\colhead{} &
\multicolumn{3}{c}{Mock Systems}\\
\cline{2-4}
\colhead{$\bm{D}$} &
\colhead{1} &
\colhead{2} &
\colhead{3}
}
\startdata
$M_{\rm 2,obs}$ ($M_{\odot}$) & 7.7$\pm$0.5 & -  & 7.84$\pm$0.25  \\
$P_{\rm orb, obs}$ (days) & - & - &  14.11$\pm$1.0 \\
$e_{\rm obs}$ & 0.69$\pm$0.05 & - &  0.47$\pm$0.05 \\
$L_{\rm x, obs}$ (erg s$^{-1}$) & - & - & 1.94$\pm$0.1$\times 10^{33}$ \\
$\alpha_{\rm obs}$ & - & 05:32:18.93 &  05:34:17.86 \\
$\delta_{\rm obs}$ & - & -70:01:02.2 &  -69:29:15.5 \\
\enddata
\end{deluxetable}

Unfortunately, this is an imperfect test since strong priors may skew the posterior distribution. Particularly in the case of HMXBs where observations may have large uncertainties and degeneracies between input parameters exist, the differences between the input values and the posterior distributions may be large. In such cases some differences, such as in the disparity between the ``observed'' orbital separation and the posterior distribution,  may be more representative of how typical are the randomly chosen initial parameters.

\subsection{Mock System 1}
\label{sec:mock_1}

Despite the caveat described above, the distribution of posterior samples and their covariances, plotted in Figure \ref{fig:mock_1}, show that our model is able to recover all the initial parameters forming our first mock system. The characteristic arc in the covariance plot comparing the posterior distributions over $a$ and $e$ from smaller separations and eccentricities to larger separations and eccentricities is due to mass transfer physics. In \bse\ binaries are assumed to instantly circularize at the pericenter separation as soon as mass transfer due to Roche lobe overflow begins. Binaries at larger separations, but lower eccentricities never reach close enough separations at pericenter to transfer mass through Roche lobe overflow, whereas binaries at larger eccentricities and shorter separations will merge.

Also of note, $a_{\rm i}$ and $e_{\rm i}$ show some evidence of bimodality in the posterior distributions, with one peak at relatively shorter separations and lower eccentricities and another at larger separations and higher eccentricities. This bimodality is derived from a bifurcation in the formation of HMXBs. Systems with initially larger separations do not transfer mass until the primary has evolved into an asymptotic giant branch star, whereas systems at shorter separations begin to transfer mass earlier in its evolution, once the donor becomes a giant star. This dichotomy combined with a correlation between $a_{\rm i}$ and $e_{\rm i}$ drives the bimodality in the joint covariance distribution of these two parameters in Figure \ref{fig:mock_1}.

The distribution of posterior samples of $\phi_{\rm k}$ in Figure \ref{fig:mock_1} are bimodal due to our formulation of SN orbital dynamics as described in Appendix \ref{sec:supernova}. Since the only dependence on $\phi_{\rm k}$ enters from a $\sin^2 \phi_{\rm k}$ term and $\sin^2 \phi_{\rm k} =\sin^2 (\pi - \phi_{\rm k})$, the distribution of posterior samples of $\phi_{\rm k}$ shows a reflective symmetry around $\phi_{\rm k} = \pi/2$ (Note that the prior on $\phi_{\rm k}$ is flat). So long as the systems of interest include a supernova kick, this symmetry can serve as an independent posterior check on any model run: The reflective symmetry around $\phi_{\rm k} = \pi/2$ should always appear in the distribution of posterior samples.

\subsection{Mock System 2}
\label{sec:mock_2}

\begin{figure*}
\begin{center}
\includegraphics[width=0.85\textwidth]{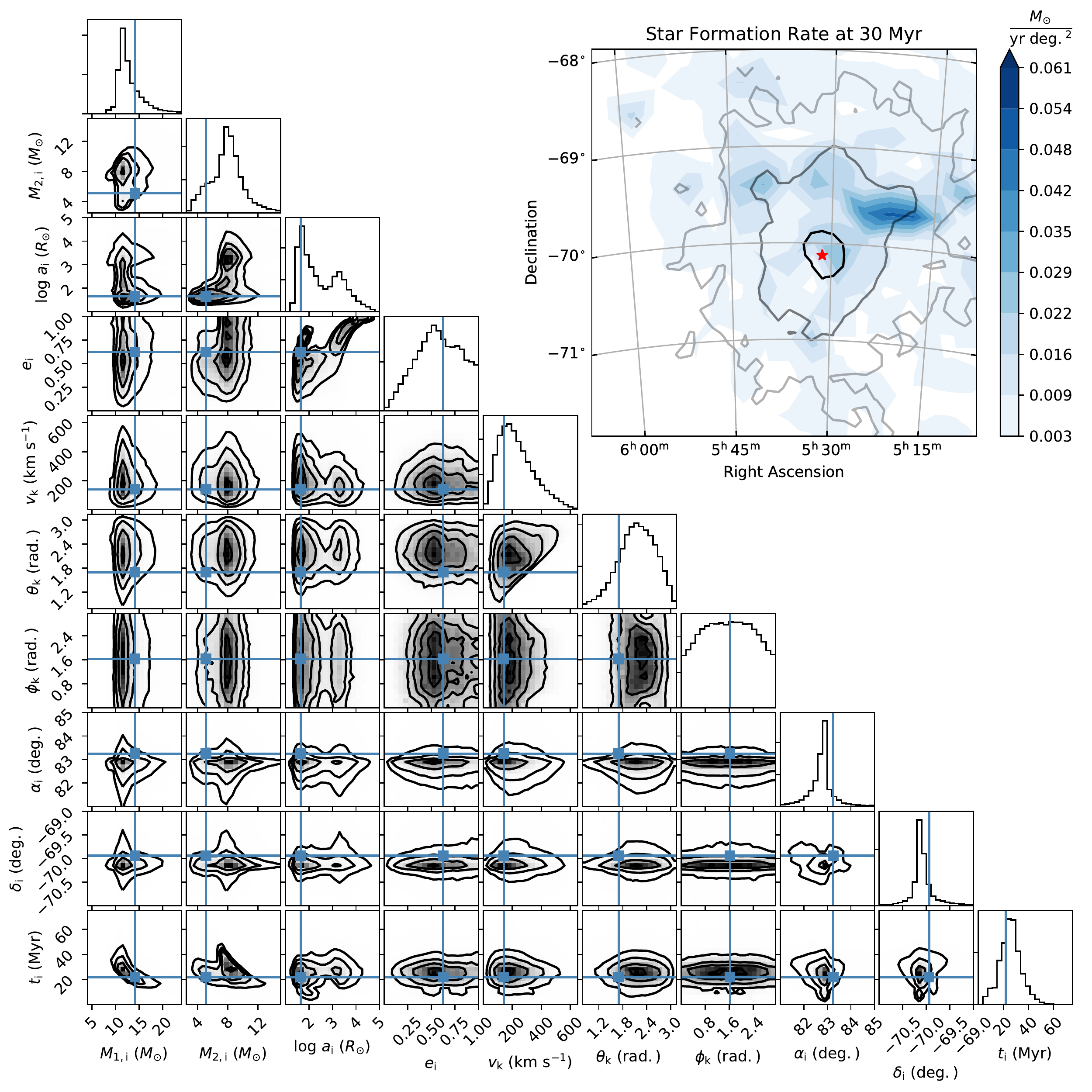}
\caption{ The covariances and 1D histograms of our posterior distribution for mock system 2. There are ten parameters in this model since we use spatially resolved SFHs and include the birth position as model parameters. In this case, the input kick velocity is at the edge of the posterior distribution. We do not expect to recover the exact birth location since there is a degree of randomness corresponding to the direction by which the system traveled as a result of the supernova kick. Nevertheless, the correct birth location is within the contours of posterior samples. The top right panel compares the posterior distribution of birth locations (greyscale contours) with the local star formation rate map (blue backgrounds) at 30 Myr, roughly the peak in the posterior distribution of birth times. Greyscale lines contain 68\%, 95\%, and 99\% of the samples, with the darkest lines indicating maxima. The current location of the mock binary is indicated by the red star. }
\label{fig:mock_2}
\end{center}
\end{figure*}

For many HMXBs in external galaxies, we may have minimal information about the orbital parameters of the system and no optical counterpart to constrain its donor. To test such a scenario, we place no constraints on the formation of mock system 2 except its observed location. Since we are now including the system's position, we have ten model parameters rather than eight as in mock system 1. More than previously, this tests how typical our choice of initial binary parameters for this system are, since we include no direct observations of the current orbital parameters.

Figure \ref{fig:mock_2} shows the posterior distribution of parameters for mock system 2. Even with minimal information about a particular system, \dart\ can identify the possible regions of parameter space forming that system by producing well sampled and defined contours. As with our first mock system, there is some evidence for multiple evolutionary scenarios which can be seen in the panel showing the covariance between $a_{\rm i}$ and $e_{\rm i}$.

There is a degree of randomness in the observed position of a particular system since it is not known which direction the supernova kick has pushed the system (or equivalently the orientation of the binary orbit in space). Nevertheless, the true birth position is near the center of the contours of the posterior distribution, which can be seen from the panel comparing $\alpha_{\rm i}$ with $\delta_{\rm i}$. The panel at the top right of Figure \ref{fig:mock_2} compares the birth position distribution to the star formation rate at 30 Myr, the peak of the posterior distribution of birth times. It can be seen that deviations in the black contours from circularity are due to spatial variations in star formation; the widest two contours, show a push to smaller right ascensions, in the direction of a region of high star formation. However, despite this rather strong prior, the posterior distribution is very much consistent with the input value.

\subsection{Mock System 3}
\label{sec:mock_3}

\begin{figure*}
\begin{center}
\includegraphics[width=0.85\textwidth]{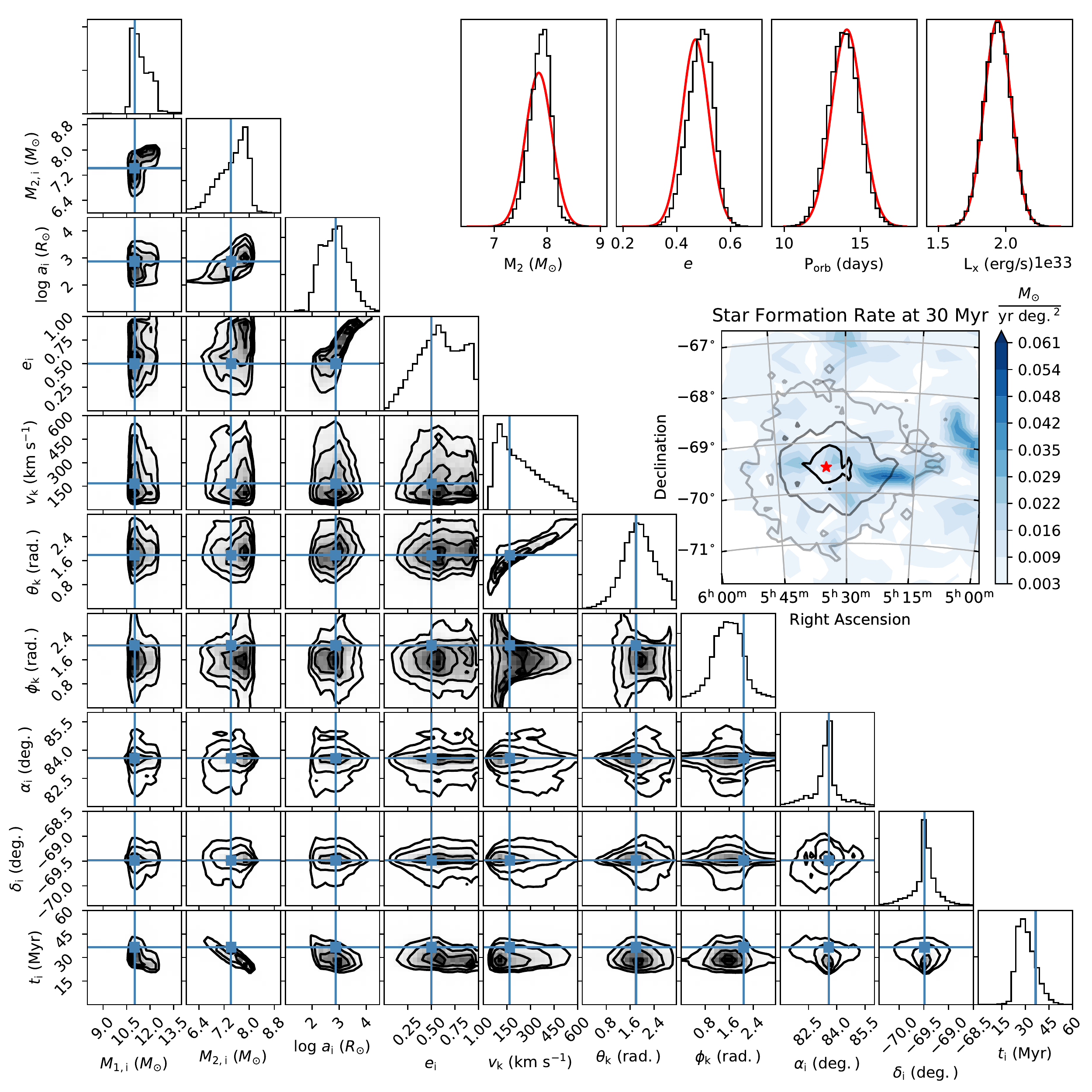}
\caption{ The covariances and 1D histograms of our posterior distribution for our third mock system. There are ten parameters in this model since we use spatially resolved SFHs and include the birth position as model parameters. As in our first and second mock examples, we are able to recover the input parameters forming this system. The four panels at the top right compare the posterior distribution of current binary parameters (black) to the four observables (red), as in Figure \ref{fig:mock_1}. The right panel compares the distribution of birth locations (greyscale contours) with the star formation rate map at 30 Myr, as in Figure \ref{fig:mock_2}. The current position of the binary is indicated by the red star. }
\label{fig:mock_3}
\end{center}
\end{figure*}

Our third mock system tests the ability of \dart\ to recover the parameters forming a binary in which the observations are precise. Included in the likelihood function for this system are constraints on the donor mass of the system, the orbital period, eccentricity, X-ray luminosity, and current position. Combined, these observations allow one to substantially reduce the possible parameter space forming a particular system.

When applying our model to the synthetic observations characterizing our third mock system, we find the likelihood function is restrictive enough that the MCMC walkers do not successfully move between separate evolutionary channels. They tend to be restricted to either a short orbital separation evolutionary channel or a large orbital separation evolutionary channel, without crossing between the two. We solve this problem by using the parallel tempering algorithm within \emcee. This method works by running multiple ensembles of walkers, each with a different ``temperature'' corresponding to a different weight applied to the likelihood function. The highest temperature ensemble has only a small contribution from the likelihood function and is therefore able to move between separate maxima efficiently. Occasionally, the walkers are allowed to switch between the parallel ensembles, causing walkers in the lowest temperature ensemble, from which the results throughout this section are derived, to fully populate the global posterior space. For more details about this algorithm and how it is applied within \emcee, see \citet{vousden16}. While this method is much better at moving around the posterior space, it is an order of magnitude more computationally expensive, since we simultaneously simulate 10 walker ensembles in our simulation, as opposed to the single ensemble of walkers we simulate elsewhere in this work.

Figure \ref{fig:mock_3} shows the distribution of posterior samples from the lowest temperature ensemble for our model of mock system 3. Most parameters are accurately recovered and precisely constrained; $M_{\rm 1,i}$ lies between 11 and 12 \Msun, and $M_{\rm 2,i}$ lies between 7 and 8 \Msun. The birth time is somewhat skewed toward shorter ages compared with the input value, and the posterior distribution of initial eccentricities is somewhat higher than expected. Nevertheless, our model recovers the input parameters fairly accurately.

The enlarged panel at the middle right of Figure \ref{fig:mock_3} compares the posterior distribution of birth locations to the local SFH at 30 Myr, near the peak of the posterior distribution of birth times. The contours spread toward the West, caused by the increased star formation in that region.

The four panels in the top right of Figure \ref{fig:mock_3} compare the posterior distributions of the observables, $M_2$, $e$, $P_{\rm orb}$, and $L_{\rm x}$, to the observational constraints on these systems (red Gaussian curves). In general, it is not expected that the posterior distributions will match the observational constraints since the posterior distribution need not match the likelihood distribution exactly. In this respect, the small inconsistency between the measured and posterior distribution of $M_2$ is expected. In this case, a combination of binary prior distributions and local SFHs has provided an improved estimate for $M_2$. Nevertheless, there is enough flexibility and degeneracy in binary evolution that the posterior probabilities of $e$, $P_{\rm orb}$, and L$_{\rm x}$ are determined almost entirely by the likelihood function. These three parameters ($e$, $P_{\rm orb}$, and $L_{\rm x}$) show excellent agreement.

\section{Application to Populations: HMXBs} 
\label{sec:results_population}

\begin{figure*}
\begin{center}
\includegraphics[width=0.85\textwidth]{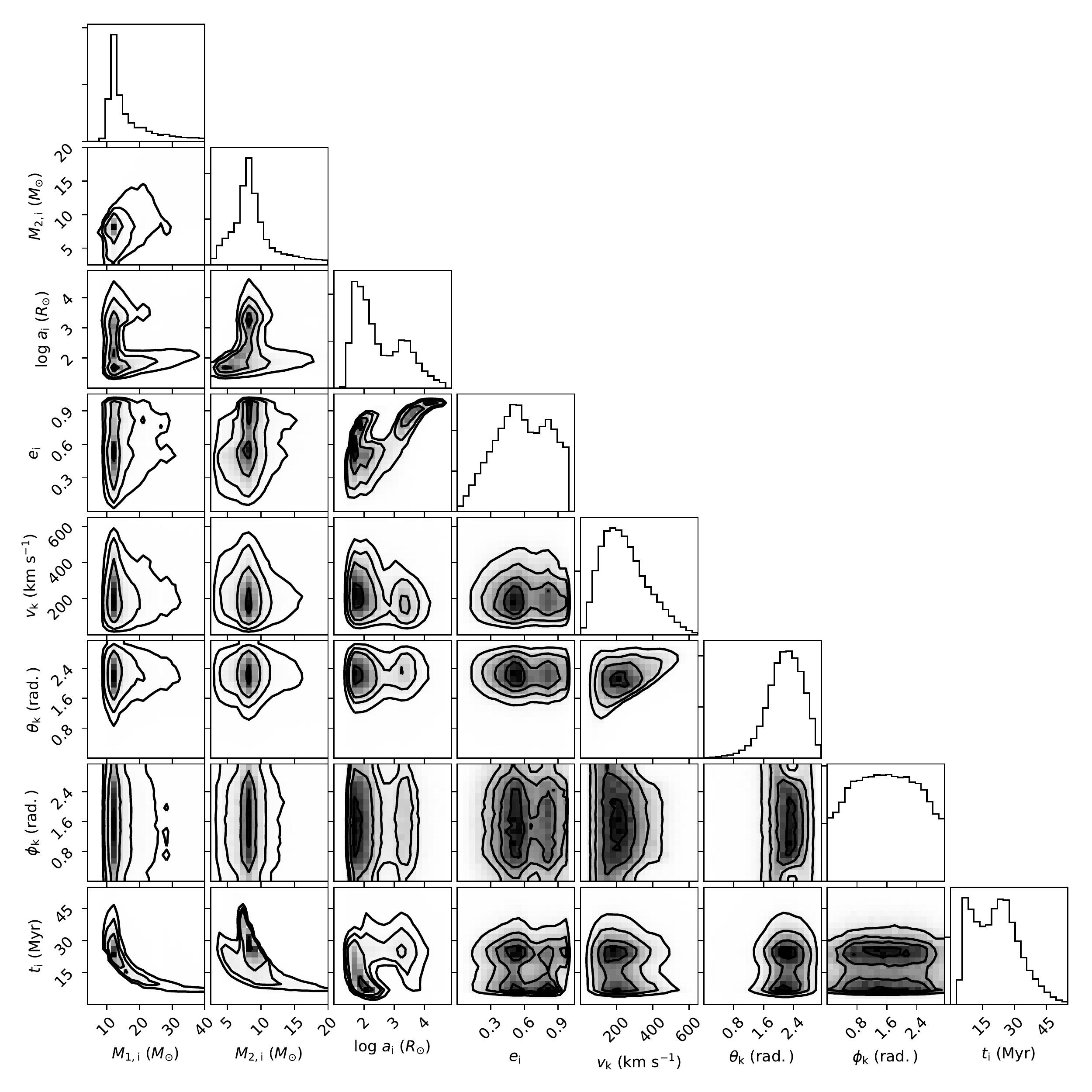}
\caption{ The covariances and 1D histograms of our posterior distribution that produces HMXBs. Some of the posterior distributions, such as $e$ and $\phi_{\rm k}$, closely reflect the priors probabilities, however others, such as $v_{\rm k}$, $\theta_{\rm k}$, and $t_{\rm i}$, show substantial differences. }
\label{fig:HMXB_corner}
\end{center}
\end{figure*}

Having tested \dart\ by recovering the input parameters producing three separate mock systems, we apply \dart\ first to the population of HMXBs. Rather than compare with data, \dart\ can give us the initial binary parameters that produce a population of binaries with certain characteristics (HMXBs in our example) expected from our understanding of binary evolution and a star formation scenario. We use the model described in Section \ref{sec:stats_population} with $x_{\rm type} = x_{\rm HMXB}$ in Equation \ref{eq:x_type} and a flat SFH. As previously stated, our definition of an HMXB is a bound system in which a NS or BH is accreting from a non-degenerate companion with a mass above 6 \Msun. No constraints are placed on the minimum X-ray luminosity (nor is there a minimum mass accretion rate) or the SFH (i.e., a constant star formation rate) since we are interested in the overall HMXB population. The posterior probabilities for model parameters that can produce such a population and their covariances are shown in Figure \ref{fig:HMXB_corner}.

Distributions over the initial masses, $M_{\rm 1,i}$ and $M_{\rm 2,i}$, strongly favor lower mass stars due to the strong weighting of the IMF toward lower masses. These indicate that according to the model chosen here, most HMXBs were formed from 10+8\Msun\ binaries, although a population exists with initial secondary masses $\approx$ 5\Msun.

The orbital separation distribution shows a dichotomy, with a high density at shorter separations and a tail extending to several thousand \Rsun. This is due to a bifurcation in the evolutionary channels forming HMXBs. The systems with smaller initial separations all overfill their Roche lobes while the primary is on the giant branch. Systems with larger initial separations avoid mass transfer at this evolutionary phase, instead overfilling their Roche lobes on the asymptotic giant branch\footnote{Both of these evolutionary channels may lead to either wind-fed or semi-detached HMXBs; the evolutionary bifurcation occurs when the primary evolves off the main sequence, before the primary undergoes core collapse.}. Orbits with very large separations will never become HMXBs since they will neither overfill their Roche lobes, nor go through a common envelope phase which is necessary to shrink the orbit.

For initial eccentricities below 0.5, the eccentricity posterior distribution reflects the prior, which linearly weights larger eccentricities. At larger eccentricities, the posterior distribution shows two distinct high probability regions, which we attribute to the same two evolutionary channels leading to a dichotomy in $a_{\rm i}$.

\begin{figure}
\begin{center}
\includegraphics[width=0.8\columnwidth]{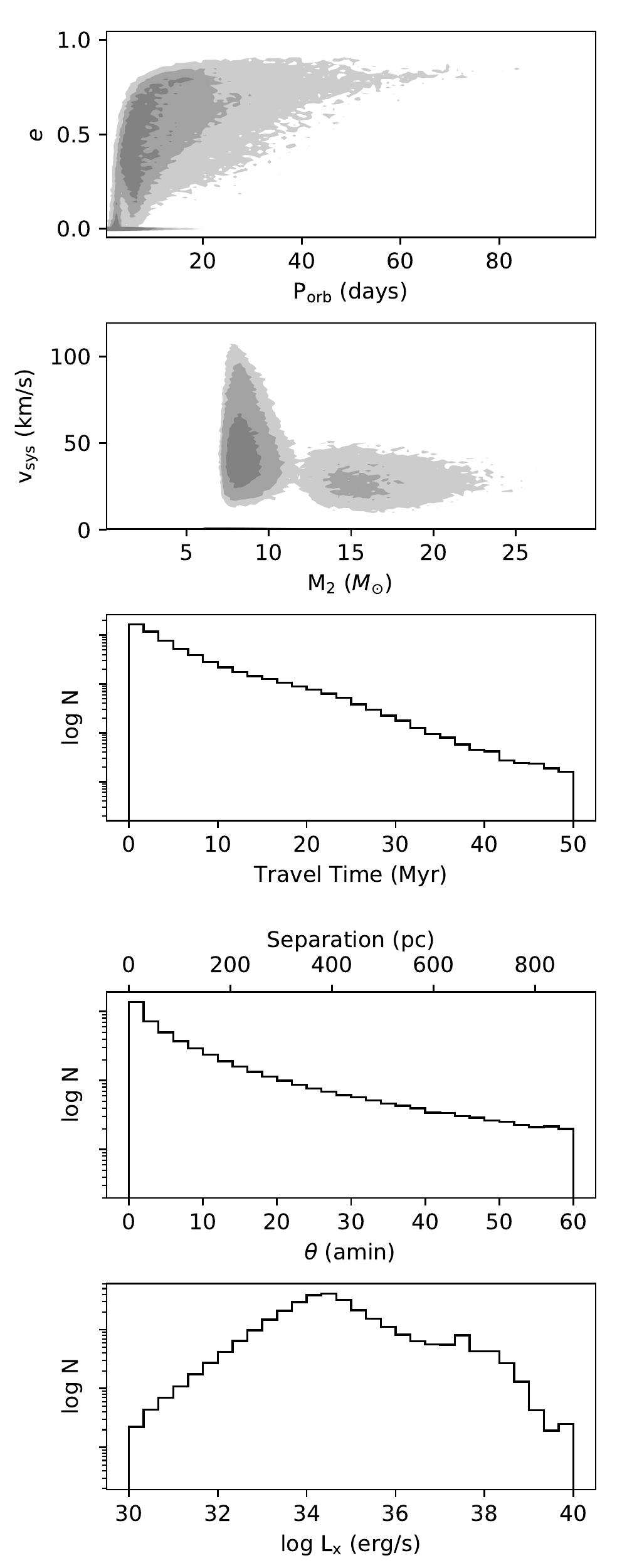}
\caption{ The present-day properties of the simulated HMXB population. The top panel shows the distribution of orbital periods, while the second panel compares the donor mass with the systemic velocity of the system. The third panel demonstrates that HMXBs are young, typically less than 10 Myr, but with a tail out to several tens of Myr. These systems therefore typically travel no farther than 10\amin, but a fraction will travel out to and beyond 1$^{\circ}$ from their birth locations. The bottom panel shows the luminosity function of HMXBs predicted by the model. }
\label{fig:HMXB_derived}
\end{center}
\end{figure}

The SN kick magnitude posterior peaks at a velocity $\approx$200 km s$^{-1}$, somewhat lower than the input Maxwellian prior (which peaks at $\sqrt{2}\sigma \approx 375$ km s$^{-1}$ for $\sigma=265$ km s$^{-1}$). The posterior distribution over the azimuthal kick angle, $\phi_{\rm k}$, is nearly flat, mirroring the flat prior on this parameter; in our orbital evolution parameterization due to SN kicks provided in Appendix \ref{sec:supernova}, $\phi_{\rm k}$ only affects the post-SN orbital eccentricity. There is a deviation from a flat distribution for $\phi_{\rm k}\approx0$, as some fraction of these systems are disrupted. We also identify a large deviation from the prior distribution for the polar kick angle, $\theta_{\rm k}$. The posterior peaks at $\pi/2$ indicating that most surviving binaries received a SN kick in the reverse of their orbital motion, as expected. The tail of the distribution in $\theta_{\rm k}$ extends toward small values, but systems surviving prograde kicks are exceedingly rare. The covariance between $v_{\rm k}$ and $\theta_{\rm k}$ hints that if the prior distribution on $v_{\rm k}$ were pushed toward smaller velocities, more binaries may survive SNe with prograde kicks.

The last row in Figure \ref{fig:HMXB_corner} shows distributions over the birth time. In this model, the star formation rate has a flat prior. Essentially all HMXBs are formed in the past 50 Myr with a peak around 10 Myr, corresponding to HMXBs with BH accretors, and a second peak in the distribution around 25 Myr, corresponding to HMXBs with NS accretors. The dominant parameters affecting this peak are the initial masses of the two stars in the binary, as seen from the panels in Figure \ref{fig:HMXB_corner} demonstrating the covariance between masses and birth times. These panels confirm our intuition that, since higher mass stars have shorter lifetimes, more recently formed HMXBs are typically formed from more massive stars, and often host BH accretors.

By taking the posterior distribution of the model parameters and evolving the binaries forward, we can determine the distribution of binary parameters today.\footnote{In evaluating the likelihood function, \dart\ already calculates each system's evolution. We use the {\tt blobs} functionality within \emcee\ to port binary parameters provided by \bse, including the system's current characteristics, to \dart. These need not be calculated a second time from the posterior samples in a post-processing step.} The top panel of Figure \ref{fig:HMXB_derived} shows the distribution of current binary orbital periods, $P_{\rm orb}$, and eccentricities, $e$; the bulk of systems have $P_{\rm orb}$ ranging from weeks to years and $e>0.5$, in broad agreement with the observed distribution of HMXB samples \citep{rajoelimanana11, liu06}. It can also be seen in the top panel of Figure \ref{fig:HMXB_derived} that a small subset of HMXBs are Roche-lobe overflowing systems which have $P_{\rm orb}$ of a few days and $e=0$.

The second panel shows the distribution of current companion (donor) masses in HMXBs, compared with systemic velocities. The most common binaries have $v_{\rm sys}\ \approx$ 40 km s$^{-1}$ and current donor star masses, $M_2$, ranging from 8-10 \Msun. This mass range is in agreement with the fact that the spectral type distribution of the donor stars in Be-type XRBs in the SMC and our Galaxy peaks at B1$-$B2 \citep{mcbride08,maravelias14}. Although note that we have used a constant star formation rate here whereas the Be-type XRB population in the SMC has been explained as a result of a peak in the star formation rate $\sim$40 Myr in the past \citep{shtykovskiy07, antoniou10}. Additionally, the systemic velocity range is consistent with constraints on systems in our Galaxy and the Magellanic Clouds \citep{vandenheuvel00, coe05, antoniou16}. A second set of systems has lower systemic velocities ($\approx$25 km s$^{-1}$) and higher companion masses ($M_2 \approx$ 15 \Msun). This subset is comprised of systems with relatively shorter initial orbital periods in which, rather than a common envelope, mass transfer proceeds on a thermal-timescale, which allows the secondary mass to substantially grow.

The third and fourth panels show the length of time a system has been traveling (the time since the primary underwent core collapse) and the angular separation between a binary's birth position and its current position. Systems may be found as much as 1$^{\circ}$ ($\approx$875 pc in the case of the LMC) away from their birth positions, but typically travel no farther than 10\amin\ ($\approx$150 pc). This result agrees with \citet{antoniou10}, who find that SMC HMXBs are typically located within regions of high star formation within the last 100 Myr, and \citet{coleiro13}, who find that Galactic HMXBs are typically separated by $\sim$100 pc from star forming complexes.

In the bottom panel of Figure \ref{fig:HMXB_derived}, we show the X-ray luminosity function produced from our posterior samples. This is calculated using the mass accretion rate onto the compact object provided by \bse\ and a conversion from gravitational energy into X-ray luminosity of 15\% for NS accretors and 40\% for BH accretors as in the prescription for wind-fed accretion by \citet{belczynski08}. The exact equation converting the mass accretion rate into $L_{\rm x}$ is provided in Section \ref{sec:L_x}. Currently, \dart\ does not calculate $L_{\rm x}$ differently for Roche-lobe overflow binaries; the $L_{\rm x}$ may be somewhat inaccurate for these systems. Our simulations show what appears to be a broken power law distribution, with a break of $\sim$10$^{34}$ \ergs\ and a second peak just above 10$^{38}$ \ergs. Since we do not account for phase-dependent mass-transfer in eccentric orbits or increased mass transfer due to Be-star HMXBs, these luminosities are only indicative.

\section{Application to Populations: HMXBs within the LMC} 
\label{sec:results_population_LMC}

\begin{figure*}
\begin{center}
\includegraphics[width=0.85\textwidth]{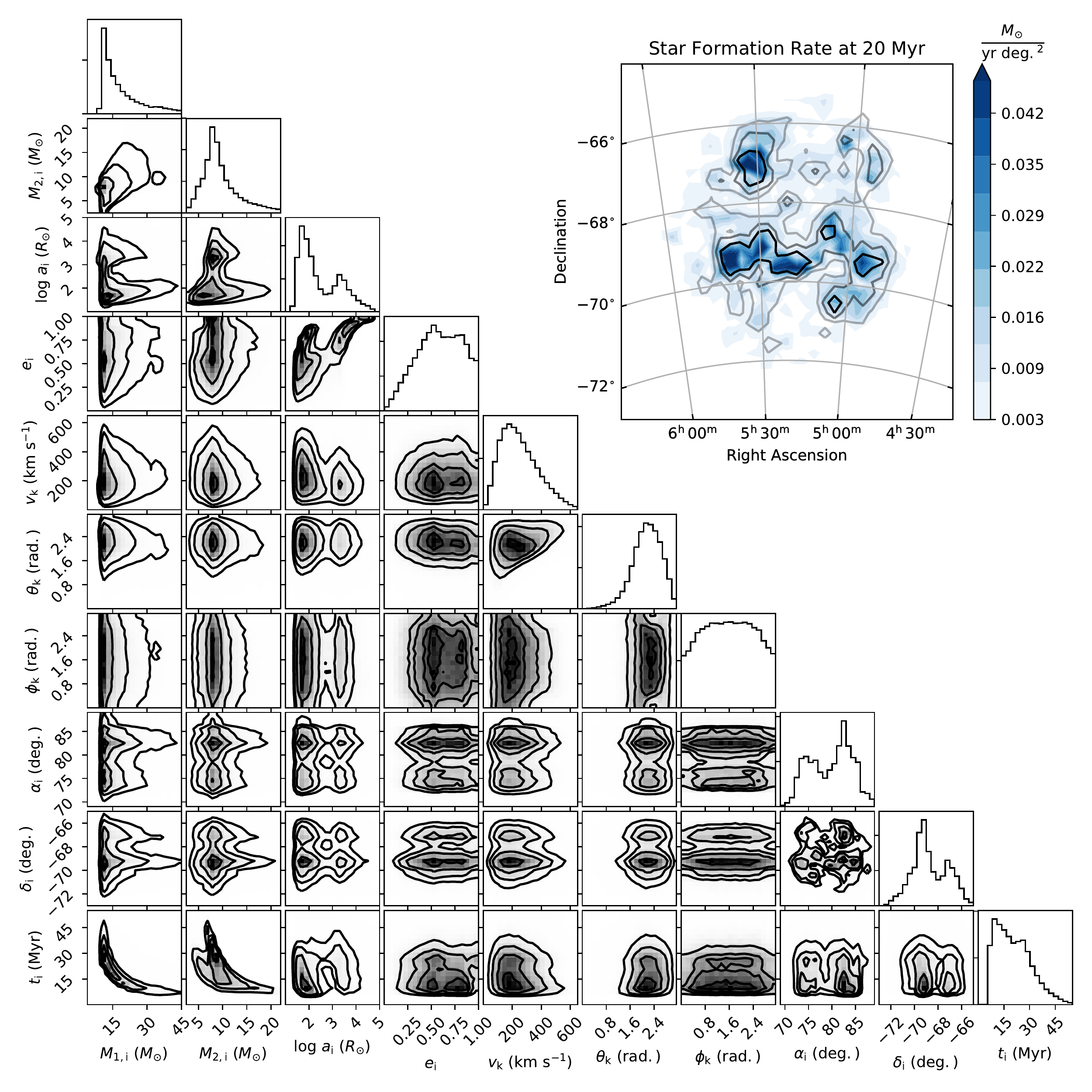}
\caption{ The covariances and 1D histograms of our posterior distribution that produces HMXBs, using the SFH of the LMC. The posterior samples produce distributions very similar to those in Figure \ref{fig:HMXB_corner}, but small differences are apparent. The posterior distribution over $t_{\rm i}$ is slightly shifted toward shorter ages, due to the recent burst of star formation within the LMC over the last 20 Myr. The panel at the top right shows the distribution of birth positions of HMXBs within the LMC. Greyscale lines contain 25\%, 50\%, and 75\% of the samples, with the darkest lines indicating maxima.  }
\label{fig:LMC_HMXB_corner}
\end{center}
\end{figure*}

In principle, one could simulate the HMXB population in the LMC (or any galaxy with a spatially resolved SFH) by post-processing the posterior sample of HMXBs discussed previously in Section \ref{sec:results_population}. Although a non-trivial exercise, this can be done by taking the already produced posterior samples and weighting each of them by the SFH of the LMC, then randomly placing them in a birth position in the LMC based on its star formation rate at that system's birth time. Weighted distributions of the posterior samples provide posterior distributions of initial binary parameters and current population parameters. However, if the star formation rate is strongly peaked at a given time (as is common in star-forming galaxies experiencing star formation episodes), a small number of simulated binaries will be weighted strongly over the others, leading to statistical uncertainty in the results that can be difficult to quantify.

Rather than performing this procedure, we adjust our model to include two extra parameters, $\alpha_{\rm i}$ and $\delta_{\rm i}$, corresponding to the right ascension and declination of the birth position of the binary. We now have a joint prior on $\alpha_{\rm i}$, $\delta_{\rm i}$ and $t_{\rm i}$ based on the spatially resolved SFH of the LMC \citep{harris09}. We prefer this method as it does not require any post-processing, can robustly deal with SFHs that deviate significantly from flat, and is generally more elegant. With a sufficiently large number of samples, the two methods should produce identical results.

Figure \ref{fig:LMC_HMXB_corner} shows the posterior sample of model parameters for our model of HMXBs within the LMC. Excluding the addition of the birth position parameters, the posterior distributions are nearly identical to those of our HMXB population model described previously in Section \ref{sec:results_population}, with one notable exception: Figure \ref{fig:LMC_SFH} shows that the LMC experienced rapid star formation in the past $\approx$20 Myr which is reflected in the excess seen in the histogram of the posterior distribution of $t_{\rm i}$ for our LMC HMXB model (bottom right panel of Figure \ref{fig:LMC_HMXB_corner}) compared with our general HMXB population model (bottom right panel of Figure \ref{fig:HMXB_corner}). Any other differences in the posterior distributions of model parameters between the two models (e.g., in $M_{\rm 1,i}$) are due to covariances between $t_{\rm i}$ and other parameters: the episodic SFH acts as a filter that allows only systems with parameters resulting in active HMXBs at the present day.

The top right panel in Figure \ref{fig:LMC_HMXB_corner} shows the distribution of birth locations in $\alpha$ and $\delta$. These are patchy which is expected since the recent star formation in the LMC is not spatially uniform \citep{harris09}. Despite the non-contiguous nature of the SFH, \emcee\ is able to explore the entire region of recent star formation demonstrating its ability to explore more generally disparate regions of parameter space.

\begin{figure}
\begin{center}
\includegraphics[width=1.0\columnwidth]{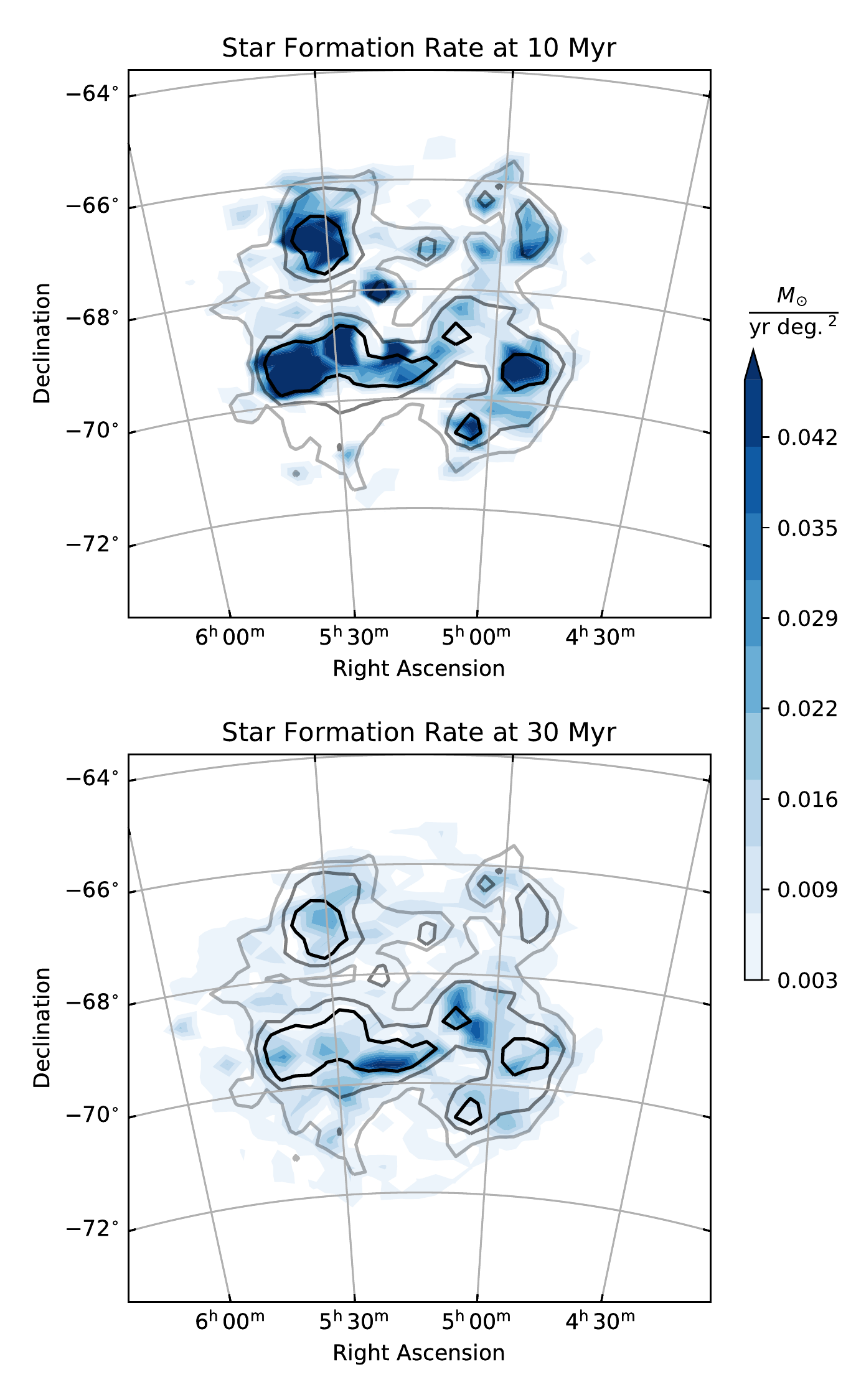}
\caption{ The expected positions of HMXBs (greyscale lines) as calculated by our model compared against the SFH (blue backgrounds) at 10 Myr (top panel) and 30 Myr (bottom panel). Greyscale lines contain 75\%, 50\%, and 25\% of the samples, with the darkest lines indicating maxima. }
\label{fig:LMC_HMXB_positions}
\end{center}
\end{figure}

By taking the posterior distribution of birth positions, determining the systemic velocity kicks these systems received and for how long they traveled, and applying a random direction to their systemic velocity, we can determine the current position distribution for HMXBs in the LMC. Figure \ref{fig:LMC_HMXB_positions} compares the model distribution of current positions (black contours) with the SFH at 10 and 30 Myr (blue background). The distribution of {\it current} locations is somewhat more extended than the star forming distribution as well as the distribution of {\it birth} locations shown in the top right panel of Figure \ref{fig:LMC_HMXB_corner}. This indicates that kick velocities can be substantial enough to move the HMXB population away from its birth position by tens of arcminutes or larger (as demonstrated by the fourth panel of Figure \ref{fig:HMXB_derived}). Note that at the LMC's distance of 50 kpc \citep{pietrzynski13}, 1 degree $\approx$ 875 pc.

In principle, one can use the posterior samples from our model to determine the distribution of current parameters of HMXBs in the LMC as was done to generate Figure \ref{fig:HMXB_derived}. However, since the distribution of model parameters are very similar, the characteristics of the current population do not substantially differ from the distributions in Figure \ref{fig:HMXB_derived}. For other galaxies, distributions of observed parameters (for instance, the luminosity function) can be extremely valuable for understanding the populations of X-ray sources \citep[e.g.,][]{tzanavaris13}. As another example, since the covariances between positions and model parameters are accounted for within our model, one can make predictions about the characteristics of systems within different parts of a specific galaxy. For instance, regions with very recent star formation may be more likely to host HMXBs with BH accretors. By selecting subregions from the posterior samples, one can quantify this likelihood. A more focused analysis of the simulation expectations of the HMXB population within the LMC and SMC will be presented in a future work.

\section{Application to Individual Systems:  {\it Swift} J0513.4-6547} 
\label{sec:J0513}

\subsection{Our Model}
\label{sec:J0513_model}

\begin{figure*}
\begin{center}
\includegraphics[width=0.85\textwidth]{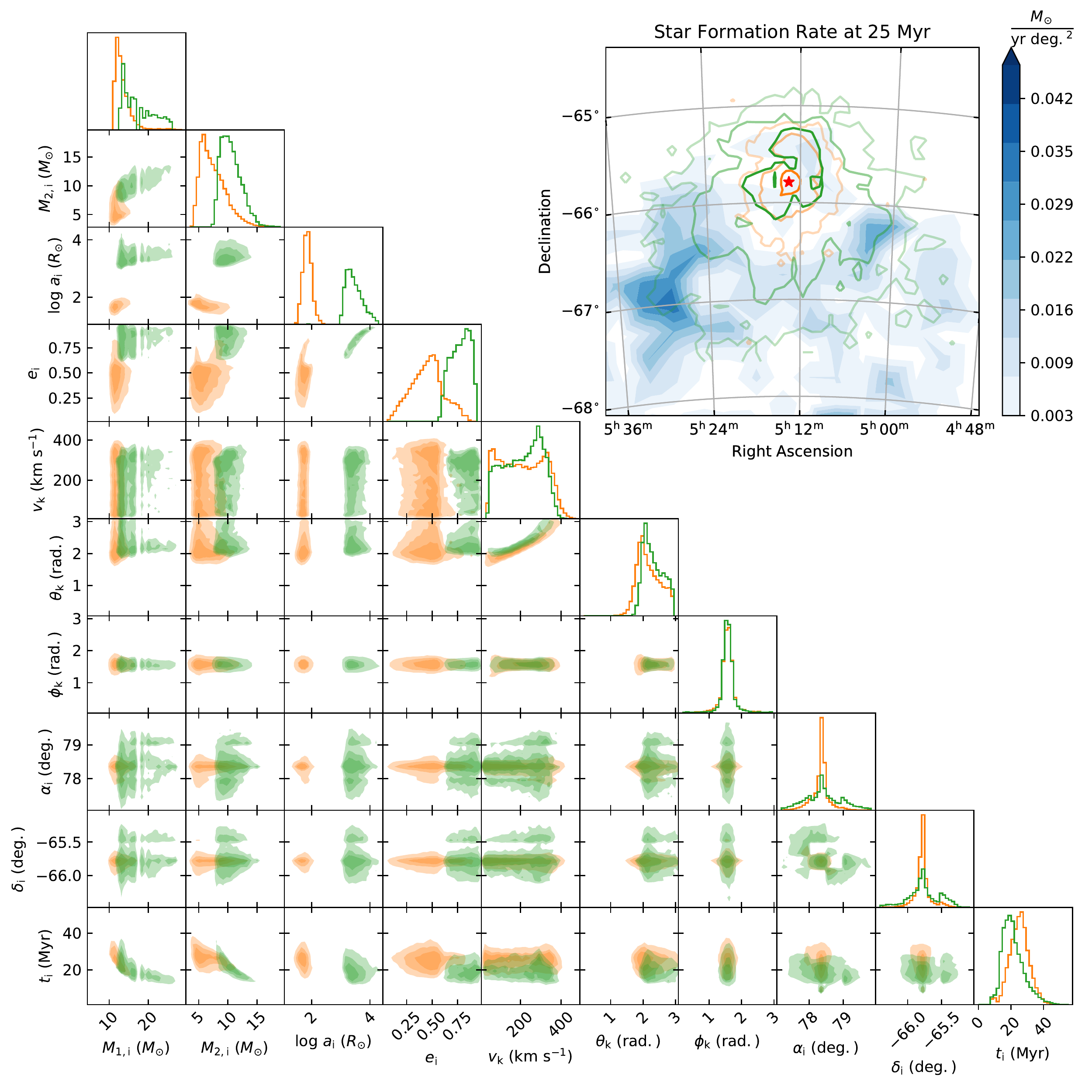}
\caption{ The 1D histograms and correlations between the ten parameters in our model for J0513. The two distributions indicate posterior samples from separate simulations that were initialized with either a relatively shorter (yellow) or {\rm longer} (green) $a_{\rm i}$. The separate distributions define separate evolutionary scenarios that both form viable solutions for the formation of J0513; however, from our calculation of the evidence ratio in Appendix \ref{sec:evidence}, the short $a_{\rm i}$ (yellow) solution is statistically preferred by a factor of $\sim$10. The top right panel shows the distribution of birth locations for J0513 (yellow and green contours, corresponding to the short and long $a_{\rm i}$ evolutionary channels, respectively) compared with the star formation rate at 30 Myr (blue background). These colored lines contain 68\%, 95\%, and 99\% of the samples, with the darkest lines indicating maxima. The red star indicates the system's current position. The long $a_{\rm i}$ evolutionary channel allows for birth locations much farther away from the present location of J0513. }
\label{fig:J0513_corner}
\end{center}
\end{figure*}

The binary system {\it Swift} J0513.4$-$6547 (hereafter J0513) was first detected by \citet{krimm09} as a pulsating X-ray source within the LMC. Analysis of X-ray and $\gamma$-ray data, as well as follow-up optical observations, allowed \citet{coe15} to identify the system as a Be-star HMXB with a B1V companion in a 27.4 day orbit. We use as input parameters the observables and their uncertainties provided by \citet{coe15} which are summarized in Table \ref{tab:J0513}. In addition to the orbital period, these include the position of the binary, an upper limit on eccentricity, and a measurement of the mass function. Although \citet{coe15} measure the orbital period to a precision of 0.008 days, we use a measurement uncertainty of 0.5 day for our model, as convergence for this exceedingly small region in posterior space can take a long time. We discuss convergence issues and possible improvements to \dart\ in Appendix \ref{sec:initialize}. We could also include the X-ray luminosity as an observable for our model to fit, however the system was discovered after it underwent an outburst. Although such outbursts are common in Be-star HMXBs, models cannot yet robustly predict their frequency, length, and luminosity for a particular system.

\begin{table}
\begin{center}
\caption{Observational constraints, taken from \citet{coe15}, included in our model of J0513. Current data only provide an orbital period, an upper limit for the eccentricity, and a mass function. The current location for the system is provided in degrees rather than sexagesimal format. Note that for computational ease we adopt a somewhat larger uncertainty on $P_{\rm orb}$ of 0.5 days rather than the value of 0.008 days as identified by \citet{coe15}. }
\label{tab:J0513}
\begin{tabular}{lc} 
\toprule
Parameter & Value \\
\midrule
$P_{\rm orb}$ & 27.405$\pm$0.5 days \\
$e_{\rm upper}$ & 0.17 \\
$m_{\rm f}$ & 9.9$\pm$2.0 \Msun \\
$\alpha$ & 78.26775 \\
$\delta$ & -65.7885278 \\
\bottomrule
\end{tabular}
\end{center}
\end{table}

Whereas with all our testing and application runs thus far, the only model that had trouble with multiple evolutionary channels was our third mock system. We were nevertheless able to successfully simulate that system using \emcee's parallel tempering algorithm. When applying our model to the observations of J0513, we found that even parallel tempering was unable to allow the walkers to move between evolutionary channels effectively.\footnote{In principle with enough separate temperature ensembles, parallel tempering should still work. Already computationally expensive, we declined an attempt at simulating J0513 with more.} The walkers tend to cluster in one of either two regions which are cleanly separated in orbital separation space.
These two regions correspond to the same evolutionary channels previously discussed: the short and long $a_{\rm i}$ solutions encompass systems in which the primary overfills its Roche lobe on the first ascent and asymptotic giant branches, respectively. After including the strict observational constraints of J0513 into the likelihood function, the valley in posterior probability between the two separate maxima is too deep and wide for the walkers to move across, even using parallel tempering with 10 separate temperatures.

To handle this difficulty, we take a separate apporach, instead running an independent simulation for each of these two evolutionary channels by initializing the walkers in a parameter space region with $a_{\rm i}$ either above or below 500 \Rsun. Throughout each simulation, the walkers remained restricted within their respective evolutionary channels. We provide further details of the initialization method we use in Appendix \ref{sec:initialize}. We find relatively longer autocorrelation lengths for this simulation, so we run both of our models for 500,000 steps, and after checking the chains for convergence, we throw away the initial burn-in of 200,000 steps. The resulting 300,000 steps provide the posterior samples from each of our two simulations.

\subsection{Model Constraints}
\label{sec:J0513_constraints}

We overplot the resulting posterior distributions of model parameters from both the short $a_{\rm i}$ (yellow) and long $a_{\rm i}$ (green) in Figure \ref{fig:J0513_corner}. While the distribution of the two separate posterior samples often overlap when projected onto many of the 2D planes, the separation between the two solutions is clear in $a_{\rm i}$, as expected. The covariance between $\alpha_{\rm i}$ and $\delta_{\rm i}$ in Figure \ref{fig:J0513_corner} demonstrates the birth coordinate distribution for J0513. In the top right panel of Figure \ref{fig:J0513_corner} we compare this distribution (yellow and green contours) to the star formation rate map (blue background) 25 Myr ago. The system (red star) currently lies in a region with little star formation at its most likely birth age. Comparison between the contours of the two simulations indicates that the long $a_{\rm i}$ solution allows for a much wider range of birth locations; however the birth location is still bounded to be within a degree or so of its current location. The posterior probability spreads toward lower declinations which have somewhat higher star formation rates. The combination of possible evolutionary channels and prior distribution of parameters such as the kick velocity disfavors the region with the highest star formation in the southeast.

\begin{figure*}
\begin{center}
\includegraphics[width=0.99\textwidth]{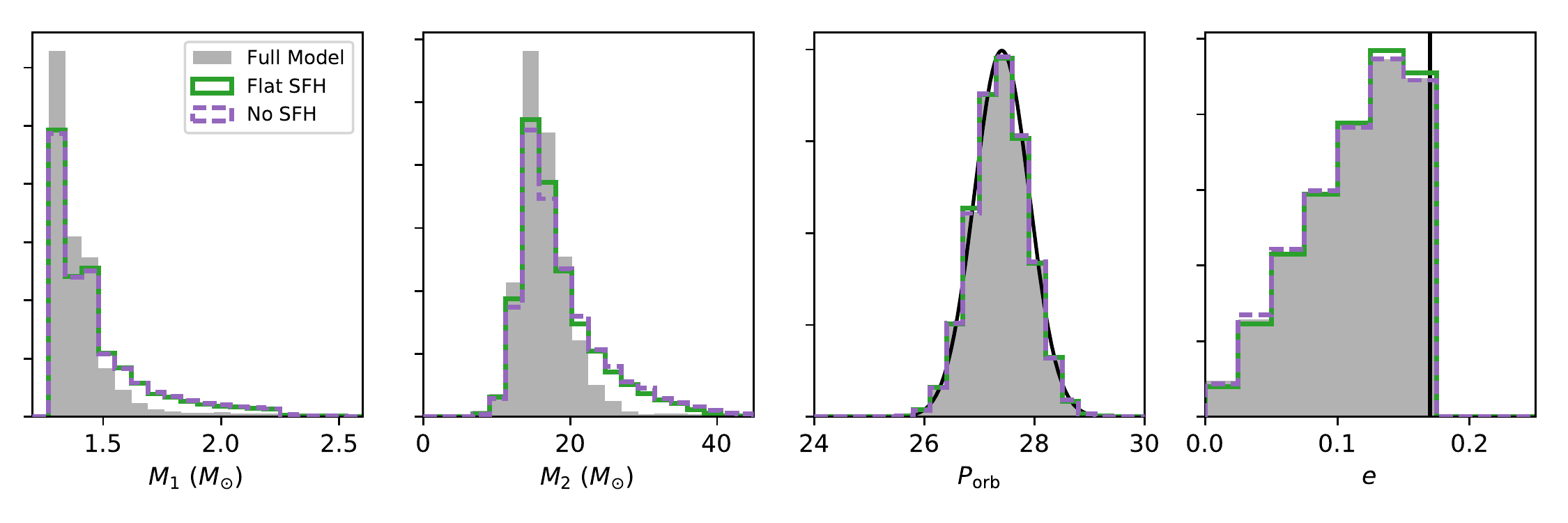}
\caption{ Model predictions for the NS mass ($M_1$), donor mass ($M_2$), $P_{\rm orb}$, and $e$ for J0513 for three different models initialized with the short $a_{\rm i}$ solution: our standard model (``Full Model''; filled, grey) as presented in Section \ref{sec:J0513_model}, one with a flat SFH throughout the LMC (``Flat SFH''; solid, green) as described in Section \ref{sec:including_SFH}, and an eight parameter model in which the position is ignored (the star formation rate is flat in this case) as described in Appendix \ref{sec:benchmarking} (``No SFH''; dashed, purple). If our model is properly constructed, the Flat SFH model and No SFH model should produce identical distributions; these distributions differ somewhat from the gray distributions depicting our Full Model, indicating that the properly applied SFH adds important information about the formation of J0513. }
\label{fig:J0513_derived_low}
\end{center}
\end{figure*}

\begin{figure*}
\begin{center}
\includegraphics[width=0.99\textwidth]{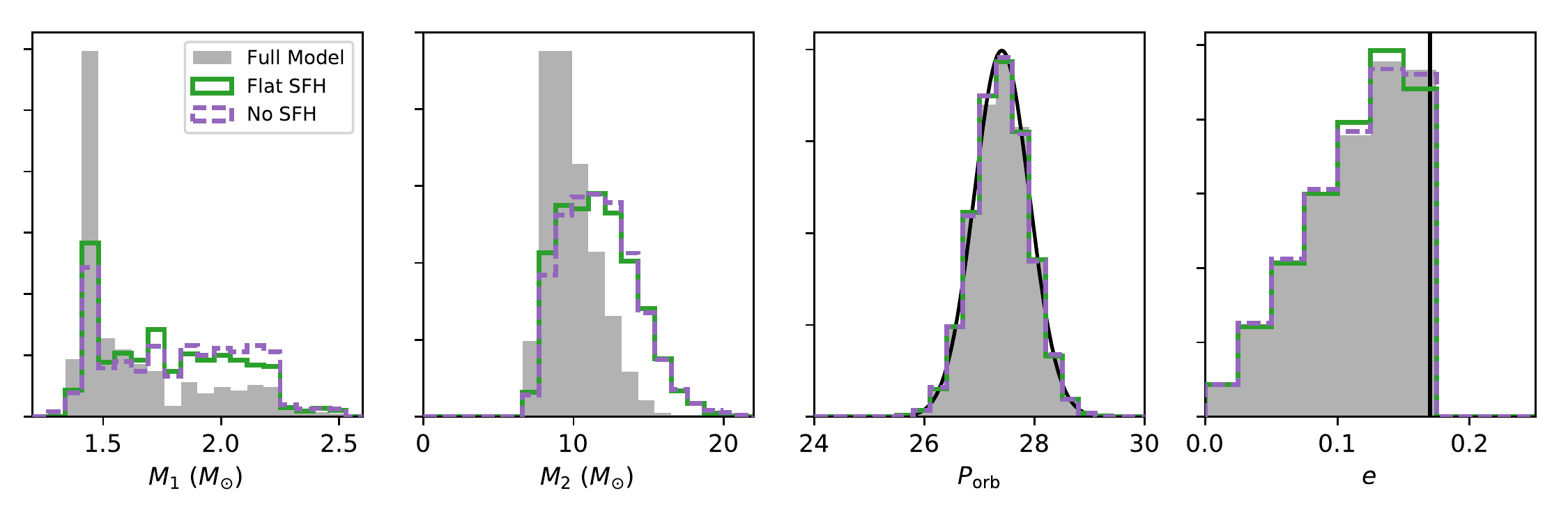}
\caption{ Model predictions for J0513 initialized with the long $a_{\rm i}$. The four parameters and three models are the same as in Figure \ref{fig:J0513_derived_low}. Here again, we see that the Flat SFH and No SFH models agree, passing our consistency check, and the Full Model produces somewhat different posterior distributions of $M_1$ and $M_2$. This again indicates that adding the spatially resolved SFH improves constraints on the system's formation. }
\label{fig:J0513_derived_high}
\end{center}
\end{figure*}

In our three mock systems, we compared the posterior distribution of model parameters to the input values. In addition to verifying the validity of our model, this comparison provides an additional check that the MCMC walkers have converged. We cannot perform such a test for J0513, since we do not know the input binary parameters forming this system in nature. As an alternative, we evolve the posterior distribution of initial binary parameters through our binary evolution prescriptions and show the posterior probability of the observables $P_{\rm orb}$, and $e$ as filled, grey distributions in Figures \ref{fig:J0513_derived_low} and \ref{fig:J0513_derived_high} for our short and long $a_{\rm i}$ evolutionary channels, respectively. The constraint on $P_{\rm orb}$ is a Gaussian (which indicates uncertainty on the measurement), while the constraint on $e$ is an upper limit. In addition to our standard model using the spatially resolved SFH of the LMC, we include two models for J0513 with different priors on star formation: a model for demonstrative purposes which has a flat, constant SFH throughout the LMC as described in Section \ref{sec:including_SFH} (``Flat SFH''; solid, green), and an eight parameter model for testing in which the position is ignored (the star formation rate is flat in this case) as described in Appendix \ref{sec:benchmarking} (``No SFH''; dashed, purple).

Assuming the model can accurately describe the observations, a converged model will probably have posterior distributions of model parameters similar to the observed values. Such a comparison provides a check on both the feasibility of the model to adequately explain the data as well as the model's convergence. In general, it is not expected that the posterior distributions exactly match the uncertainties on the observables; non-flat prior distributions on the model parameters and binary evolution in general will skew the resulting posterior distribution of observables. However, the posterior and observed distributions over $P_{\rm orb}$ are nearly identical for both evolutionary channels, regardless of the SFH model. Figures \ref{fig:J0513_derived_low} and \ref{fig:J0513_derived_high} also show the posterior distribution of the current orbital eccentricity. Although the observations only provide an upper limit, our model makes a prediction that the orbit of the system is non-circular. This is an example of how our model can improve the constraints on system characteristics, even those that are not directly observable.

Figures \ref{fig:J0513_derived_low} and \ref{fig:J0513_derived_high} further show model predictions for other current parameters of the system. In this case, we show the posterior distribution of the NS mass, $M_1$, and the donor mass, $M_2$. Our model predicts a NS mass less than 1.5 \Msun, close to the canonical value, and a companion mass of $\approx$10 or $\approx$15 \Msun, for the short and long $a_{\rm i}$ evolutionary channels, respectively. Although it is not included as part of the likelihood function, the spectral type of the donor \citep[B1V;][]{coe15} indicates a mass of $\approx$13 \Msun, consistent with both evolutionary channels.

Table \ref{tab:J0513_results} shows quantitative results from our ``Full Model'' analysis. Of course, we have two separate solutions depending on whether the walkers were initialized with a short or long $a_{\rm i}$. Determining which of the two solutions is preferred cannot be straightforwardly done with only the posterior samples from these two simulations, as it requires the calculation of the Bayesian evidence. In Appendix \ref{sec:evidence} we describe a method for this calculation, and demonstrate that the short $a_{\rm i}$ solution is preferred by a factor of $\gtrsim$10. The values listed in Table \ref{tab:J0513_results} for our ``Full Model'' using a short $a_{\rm i}$ provide our best model constraints for J0513.

\renewcommand{\arraystretch}{1.5}

\begin{table*}
\begin{center}
\caption{1-$\sigma$ confidence levels for posterior samples from our full model of J0513 (described in Section \ref{sec:J0513_model}), our test model in which the LMC is assumed to have a flat SFH (described in Section \ref{sec:including_SFH}), and our model ignoring positional information (described in Appendix \ref{sec:benchmarking}). For nearly every parameter, our full model which takes into account the position of J0513 and includes the SFH produces more precise constraints than our models ignoring that information.}
\label{tab:J0513_results}
\begin{tabular}{lccccccc} 
\toprule
 & \multicolumn{3}{c}{Short $a$ Solution} &  & \multicolumn{3}{c}{Long $a$ Solution} \\
\cline{2-4} \cline{6-8}
Parameter & Full Model & Flat SFH & No SFH & & Full Model & Flat SFH & No SFH \\
\midrule
$M_{\rm 1,i}$ (M$_{\odot}$) & 12.3$\substack{+2.5 \\ -1.3}$ & 13.0$\substack{+4.9 \\ -1.9}$ & 13.1$\substack{+5.2 \\ -2.0}$ &
    & 14.0$\substack{+5.7 \\ -1.2}$ & 16.7$\substack{+5.4 \\ -3.1}$ & 18.4$\substack{+5.3 \\ -4.7}$ \\
$M_{\rm 2,i}$ (M$_{\odot}$) & 6.7$\substack{+3.1 \\ -2.0}$ & 7.4$\substack{+4.5 \\ -2.5}$ & 7.5$\substack{+4.8 \\ -2.6}$ &
    & 9.4$\substack{+2.2 \\ -1.3}$ & 11.4$\substack{+2.7 \\ -2.5}$ & 11.6$\substack{+2.7 \\ -2.5}$ \\
$a_{\rm i}$ (R$_{\odot}$) & 52$\substack{+25 \\ -16}$ & 55$\substack{+25 \\ -16}$ & 55$\substack{+26 \\ -17}$ &
    & 2680$\substack{+3320 \\ -1140}$ & 3150$\substack{+4690 \\ -1340}$ & 3240$\substack{+4050 \\ -1350}$ \\
$e_{\rm i}$ & 0.46$\substack{+0.17 \\ -0.19}$ & 0.46$\substack{+0.19 \\ -0.20}$ & 0.47$\substack{+0.20 \\ -0.20}$ &
    & 0.81$\substack{+0.10 \\ -0.12}$ & 0.82$\substack{+0.10 \\ -0.13}$ & 0.82$\substack{+0.10 \\ -0.12}$ \\
$v_{\rm k}$ (km s$^{-1}$) & 200$\substack{+139 \\ -135}$ & 195$\substack{+144 \\ -130}$ & 196$\substack{+149 \\ -132}$ &
    & 192$\substack{+109 \\ -133}$ & 116$\substack{+145 \\ -80}$ & 145$\substack{+146 \\ -98}$ \\
$\theta_{\rm k}$ (rad.) & 2.17$\substack{+0.56 \\ -0.31}$ & 2.15$\substack{+0.51 \\ -0.28}$ & 2.15$\substack{+0.54 \\ -0.29}$ &
    & 2.32$\substack{+0.48 \\ -0.27}$ & 2.18$\substack{+0.37 \\ -0.18}$ & 2.21$\substack{+0.43 \\ -0.19}$ \\
$\phi_{\rm k}$ (rad.) & 1.57$\substack{+0.16 \\ -0.17}$ & 1.58$\substack{+0.16 \\ -0.16}$ & 1.57$\substack{+0.17 \\ -0.17}$ &
    & 1.57$\substack{+0.17 \\ -0.16}$ & 1.57$\substack{+0.19 \\ -0.21}$ & 1.57$\substack{+0.18 \\ -0.18}$ \\
$t_{\rm i}$ (Myr) & 25.1$\substack{+5.6 \\ -5.9}$ & 23.1$\substack{+7.0 \\ -8.5}$ & 22.7$\substack{+7.3 \\ -8.5}$ &
    & 23.4$\substack{+7.5 \\ -6.1}$ & 16.9$\substack{+7.3 \\ -4.2}$ & 16.2$\substack{+7.6 \\ -4.6}$ \\
\bottomrule
\end{tabular}
\end{center}
\end{table*}

\renewcommand{\arraystretch}{1.0}

\subsection{Why Include Star Formation Histories?}
\label{sec:including_SFH}

\begin{figure*}
\begin{center}
\includegraphics[width=0.85\textwidth]{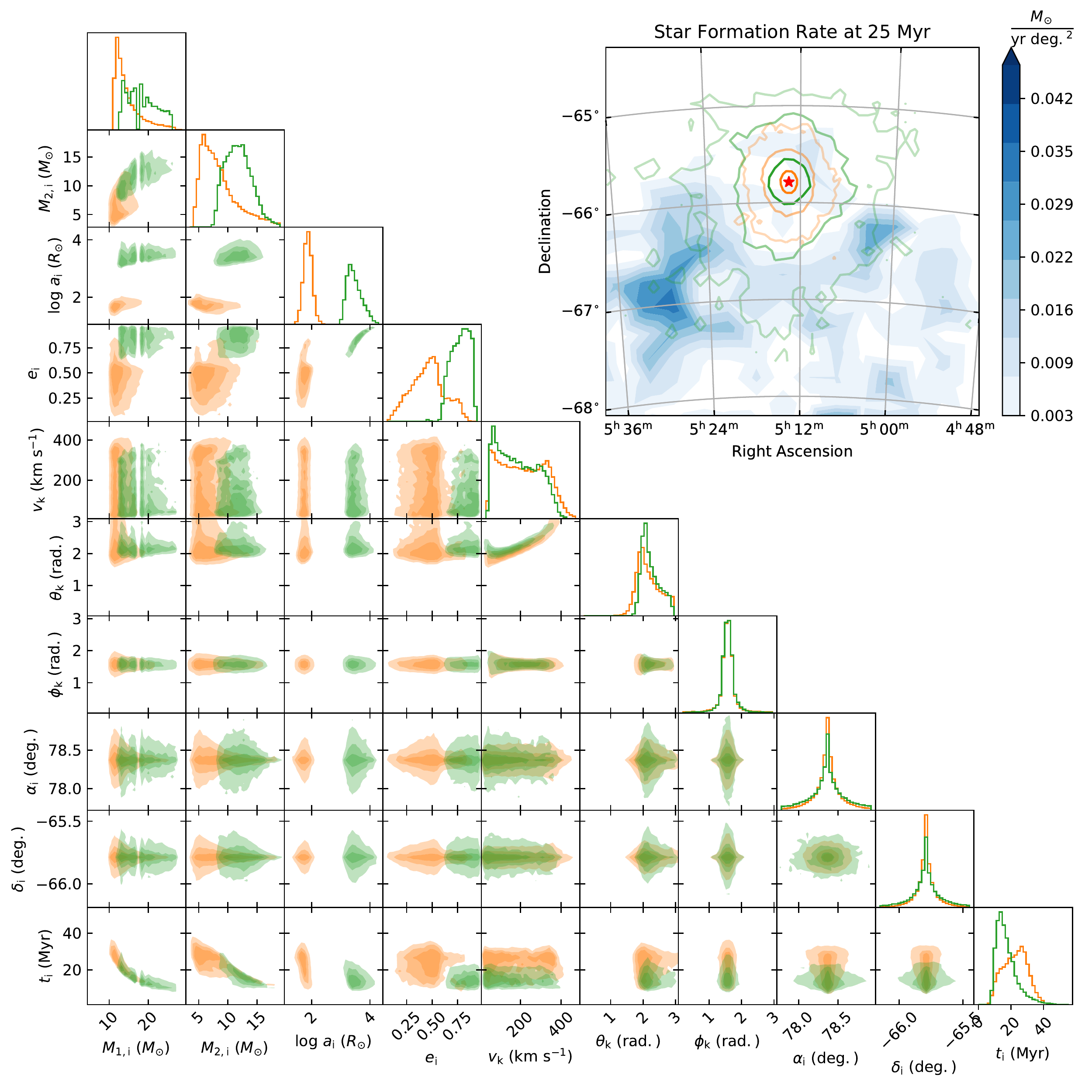}
\caption{ The 1D histograms and correlations between the ten parameters in our model for J0513, using a flat SFH for LMC, and either the short $a_{\rm i}$ or long $a_{\rm i}$ solution. As in Figure \ref{fig:J0513_corner} the top right panel shows the distributions of birth locations for J0513 (colored contours). For reference, we compared these distributions with the star formation rate at 30 Myr. The red star indicates the system's current position. Both evolutionary channels produce rotationally symmetric posterior distributions in the birth location surrounding the system's position, as expected for a flat SFH prior. }
\label{fig:J0513_flatsfh_corner}
\end{center}
\end{figure*}

J0513 resides in a position in the LMC in which the most recent star formation episode occurred around 25 Myr ago. This additional information constrains the formation path of J0513. As a demonstration of this, we run an additional model of the formation of J0513, as was done in Section \ref{sec:J0513_model}, except using a ``flat'' SFH for the LMC that is constant in time and space. Figure \ref{fig:J0513_flatsfh_corner} shows the distribution of posterior samples of our model using this demonstrative SFH.

The clearest difference between the two posterior samples can be seen in the bottom right-most panel showing the 1D histogram of birth times. This model, with a flat SFH, has a posterior distribution of birth times skewed toward younger ages compared to that of our model with the actual SFH of the LMC. The younger ages result in relatively more massive initial masses for the binary's components. These are the most significant differences between the two models, but other minor variations can be seen. For instance, the birth location distribution is now axisymmetric, which can be seen from the panel at the top right of Figure \ref{fig:J0513_flatsfh_corner}, as expected from a non-spatially varying SFH. Furthermore, as expected, with the inclusion of one additional constraint (the SFH information), the model parameters are much better constrained.

For J0513, one may be able to obtain reasonable results by running a simplified model without the birth position as a parameter, instead applying a prior to the birth time which corresponds to the SFH of the closest pixel in the SFH map. However, the SFH maps of the LMC have a somewhat finer resolution than the typical distances traveled by a HMXB. Indeed, this is why many HMXBs are found some distance away from the star forming regions (assuming the SFH maps have a fine enough resolution) in which they were presumably born \citep[see][although note that this result is for more distant galaxies.]{kaaret04}. The resolution with which SFHs can be probed may be coarser than typical distances traveled by HMXBs for more distant galaxies. In such cases, a simplified model using merely the SFH of the closest pixel may provide an accurate result. However, for nearby galaxies, this is not the case. As methods of calculating the spatially resolved SFH improve and the angular precision with which such histories can be determined becomes smaller, inclusion of this information becomes more important. As Figure \ref{fig:J0513_corner} demonstrates, with precise SFH maps, the stellar cluster or star forming region where a particular system was born can be identified which in turn allows for constraints on the SN kick velocity.

The results in Table \ref{tab:J0513_results} provide a quantitative example of the importance of including spatially resolved SFHs. In this table, we provide the median value of the input parameter, with uncertainties corresponding to the 68\% confidence interval on the median. Comparison of the uncertainties on parameters between the ``Full Model'' described in Section \ref{sec:J0513_model} and the ``Flat Model'' described in this section shows that, at least in this example, the SFH can substantially improve the constraints on nearly every parameter defining a system's formation.

\section{Discussion and Conclusions}
\label{sec:discussion}

\subsection{A New Approach in BPS}

Past observations have shown that the population of extragalactic HMXBs are often found near, but not necessarily coincident with, regions of high star formation \citep{zezas02b}. \citet{sepinsky05} showed that SN kicks can lead to a displacement between a binary's current position and its birth location. At the same time, the SN kick affects the binary's orbit, and a correlation should exist between binary parameters and the distance an HMXB travels from its birth site \citep{zuo10,zuo15}. These models use traditional BPS to reproduce general characteristics of the HMXB population, such as the observation that HMXBs with higher X-ray luminosities tend to be found closer to star forming regions than systems with lower X-ray luminosities \citep{kaaret04}.

Traditional BPS is too inefficient to correlate the local SFH with binary characteristics for individual systems; too many of the Monte Carlo generated systems merge or disrupt at some point during their evolution. Of those that do evolve into HMXBs only a fraction evolve into systems of interest for a given study. The best-observed HMXBs, those systems with the most potential to constrain binary evolution, are often the least efficient to model since many separate observational characteristics need to be simultaneously matched. The situation is even more difficult for LMXBs, which undergo more complex evolutionary paths \citep{kalogera96b, kalogera98, posdiadlowski02}.

In this work we describe a fully Bayesian method that allows for more detailed comparisons between HMXBs and SFHs. We interpret BPS as a parameter estimation problem that includes the spatially resolved SFH as a prior, together with observations of individual systems and their uncertainties in the likelihood function. Prior probabilities for the binary parameters and SN kicks are based on the same distributions used by traditional BPS.

\dart\ includes several novel features: First, simultaneous consideration of source position and orbital parameters, which can set more stringent constraints on the formation of systems. Second, an MCMC fitting approach to explore the parameter space, which provides more efficient sampling, particularly important for rare or short-lived systems. Third, flexible inclusion of different observational constraints or biases (e.g., incompleteness) without the need for fine tuning of the methodology, allowing for a consistent approach to a heterogeneous data set. Finally, adaptability to study both individual systems as well as populations of systems.

Our approach avoids some of the common problems of traditional BPS. When comparing to an observed populations, traditional BPS studies often ignore observational uncertainties (or treat them in an ad hoc manner); whereas uncertainties are seamlessly included in our Bayesian formalism in the form of a likelihood function. Unique or rare systems may have been formed in relatively low probability regions of the parameter space, and using traditional BPS, it is both difficult to synthesize a statistically substantial number of systems and to know if those systems fully represent the parameter space. By moving through the parameter space based on the {\it posterior} probability, rather than making random draws from the prior probabilities, our MCMC model is able to efficiently generate a statistically significant distribution of posterior samples. This efficiency translates into the faster inclusion of updates to binary evolution physics and a faster comparison between XRB evolution models with different physical prescriptions.

\subsection{Current Limitations and Future Directions}
\label{sec:limitations}

The potential applications for MCMC in BPS are numerous, as are the potential pitfalls. Careful attention needs to be paid to ensure the prior probabilities are properly described since binary parameter priors are often strong; the prior distributions should be properly normalized, particularly when model parameters have joint priors such as with $M_{1,{\rm i}}$ and $M_{2,{\rm i}}$ or with $\alpha_{\rm i}$, $\delta_{\rm i}$, and $t_{\rm i}$ in our model.

The posterior chains need to have converged to draw conclusions from the model. Although convergence cannot be guaranteed, we suggest two informal methods to check for convergence in the chains: First, visual inspection of the trace should indicate that the dispersion of chains does not considerably vary as the chain progresses. Second, when initializing the simulation in separate locations in parameter space, the resulting posterior samples ought to produce identical distributions to statistical variance. One should also check that the posterior distribution of observables should roughly correspond to (but not necessarily exactly mimic) the values indicated by the observations and their uncertainties.


A potential downside of MCMC methods is that they can have problems moving across sharp boundaries in parameter space. In our tests, we have seen that the chains typically are able to transition between different evolutionary channels. We have tested this by initializing the walkers in separate locations. For systems with precise observations which increase the height of the likelihood barrier between separate maxima in probability space, we have found that the model can have difficulties transitioning between two general evolutionary channels which are defined by whether the system overfills its Roche lobe on the giant branch or asymptotic giant branch. This difficulty was overcome for our third mock system using a parallel tempering algorithm \citep[e.g,][]{vousden16} which we provide as an option within \dart. For J0513, even parallel tempering was unable to deal with the multiple evolutionary channels. In this case, we applied a separate technique in which we ran separate simulations for the two evolutionary channels we identified. Afterwards, we used a method described in Appendix \ref{sec:evidence} to calculate the ratio of the Bayesian evidences for each channel, thereby determining which of the two evolutionary channels was preferred.

Other aspects of constraining binary formation based on quantitatively combining observations with SFHs may cause separate local maxima. For instance, the LMC clearly shows non-contiguous star formation. Nevertheless, the top right panel of Figure \ref{fig:LMC_HMXB_corner} shows that the walkers are successfully distributed throughout different star forming regions. This suggests that other less obvious boundaries may not be problematic. However, we note that even if \emcee\ successfully moves across different regions of parameter space, sharp boundaries and multiple maxima in the parameter space may lead to a decreased acceptance fraction.

An additional limitation of using MCMC rather than traditional importance sampling for BPS is that MCMC cannot straightforwardly provide rates; the evidence integral (denominator in Equation \ref{eq:bayes_ind}) must be calculated, which is a non-trivial exercise from a set of posterior samples. Typical methods such as thermodynamic integration are stymied by the requirement that $\bm{x}_{\rm f} \in x_{\rm type}$ which effectively denormalizes the prior distributions; thermodynamic integration could allow one to calculate the evidence ratio for a system of some class $x_{\rm type}$ satisfying some observational constraints {\it with respect to all systems of that class}, but not with respect to the entire population of stellar binaries. Note that this is not a problem for constraining BPS model parameters, determining the initial and present day parameters of a particular system, or simulating the HMXB population of a particular galaxy. Therefore we cannot calculate the absolute number of HMXBs, although we note that a method similar to the one used to calculate the ratio between the Bayesian evidences for separate evolutionary channels described in Appendix \ref{sec:evidence} can provide rates for systems with specific observations relative to the overall number of HMXBs. We defer a further exploration of this for future work.

The relative efficiency of our method compared to traditional BPS depends on the problem at hand (see Appendix \ref{sec:benchmarking} for a demonstration). Traditional methods will be less efficient when the binary population is rare or short-lived and therefore the region of relevant parameter space is smaller (such as the merging double compact object systems that produce gravitational wave radiation). However, our method suffers from inefficiencies as well: each step is related to the previous one, therefore the set of independent posterior samples is reduced by a factor of the autocorrelation length. For the cases presented in this work, we find relatively long autocorrelation lengths (see Appendix \ref{sec:initialize}). The chains must be run for enough steps that the region of viable parameter space has been thoroughly explored.

Although our method efficiently identifies the initial parameters forming a specific HMXB (or population of HMXBs), this is only the first step. The technique described in this work can be used to determine aspects of the system that are not observable directly, such as the binary's formation scenario or whether it likely hosts a NS or BH. Furthermore, the same MCMC techniques can be applied to well-measured X-ray binaries in nearby galaxies (as was done here for J0513) using different binary evolution prescriptions. Ultimately, after combining similar analyses from multiple binaries, identifying the initial parameters forming the observed population can lead to constraints on underlying models for key binary evolution physics, such as the kicks applied to systems at birth; an aggregate analysis can allow us to not only constrain the formation of individual systems, but the prior distributions for stellar binaries as well.

Our model can be expanded in a number of ways. First, we plan to expand this model to other X-ray binaries (including both HMXBs and LMXBs) in a variety of nearby galaxies including the SMC and M81, as well as the LMC as was done here, each of which also have well understood SFHs \citep{harris04, harris09} and HMXB populations \citep[e.g.,][]{antoniou16}.

With the recent LIGO detections of compact object mergers \citep{abbott16a,abbott16b, abbott17a, ligo17a, ligo17b}, there is renewed interest in deriving evolutionary histories for merging, evolved massive binaries \citep[e.g.,][]{belczynski16}. As instrument sensitivity improves and the number of detections increases, new methods for deriving evolutionary histories for merging compact objects will be crucial for fully understanding the formation of these systems and modeling their population. Although our model is derived for HMXBs, a subset of these systems, particularly those with massive Wolf-Rayet components that are unlikely to disrupt after the second component's core collapse, may evolve into binary black holes \citep{belczynski13,maccarone14,vandenheuvel17}. With updates to the binary evolution physics, our MCMC approach could provide a natural, general method by which to efficiently derive posterior distributions for those systems forming merging compact object binaries. Specifically, by using the SFH of the galaxy hosting a merging double compact object (by association of a LIGO event with an electromagnetic counterpart and a nearby host galaxy) as a prior, more stringent constraints on a system's evolutionary history can be made.

In addition to new applications for \dart, we also plan to develop new features. For instance, whereas we have demonstrated here how \dart\ can be applied to individual systems to derive their evolution, by applying \dart\ to multiple systems, one can use the combined posterior samples to constrain binary evolution prescriptions (e.g., the SN kick dispersion velocity). Alternatively, a more complex, hierarchical model could simultaneously constrain formation scenarios for the observed systems as well as the parameterizations themselves. Such a model requires a Bayesian approach such as the one described here \citep{lee11, xu14, wysocki17, zevin17}.

\subsection{Conclusions}
\label{sec:conclusions}

We describe a fully Bayesian method to identify the binary initial conditions forming both individual HMXBs and HMXB populations. Our method includes both binary evolution physics as well as spatially resolved SFHs to constrain the formation channels of HMXBs. As a test, we apply our method to three individual mock binaries, and our method is generally able to recover all the initial parameters of these binaries. Having passed this test, we apply our model to the population of HMXBs as well as the population of HMXBs within the LMC. Finally, we apply our model to the LMC HMXB {\it Swift} J0513.4$-$6547, and our model converges on the region of parameter space forming this binary; the posterior distribution of model parameters produces HMXBs matching the observations of this system. Our model is, by construction, flexible to allow the inclusion of different X-ray binaries with different observables. Furthermore, it forms the basis for future, hierarchical models that would allow us to constrain the formation and evolutionary parameters for populations of X-ray binaries in individual as well as samples of galaxies. Our model, and MCMC techniques more generally, have the potential to become a powerful tool for the study of binary populations.

\section*{Acknowledgements}
We thank the anonymous referee for helpful suggestions that greatly improved the manuscript. We thank Vicky Kalogera, Richard O'Shaunessy, Stephen Justham, Ilya Mandel, and Will Farr for useful conversations. J.J.A. and A.Z. acknowledge funding from the European Research Council under the European Union's Seventh Framework Programme (FP/2007-2013)/ERC Grant Agreement n. 617001. T.F. acknowledges support from the Ambizione Fellowship of the Swiss National Science Foundation (grant PZ00P2-148123). This project has received funding from the European Union's Horizon 2020 research and innovation programme under the Marie Sklodowska-Curie RISE action, grant agreement No 691164 (ASTROSTAT). This work was performed in part at Aspen Center for Physics, which is supported by National Science Foundation grant PHY-1607611. This work was partially supported by a grant from the Simons Foundation. The authors acknowledge the International Space Science Institute (ISSI) for supporting and funding the international teams program ``Ultraluminous X-ray Sources: from the Local Group to the Very First Galaxies.'' We acknowledge use of the Metropolis HPC Facility at the CCQCN Center of the University of Crete, supported by the European Union Seventh Framework Programme (FP7-REGPOT-2012-2013-1) under grant agreement no. 316165.

\software{astropy \citep{astropy}, corner \citep{corner}, dart\_board \citep{dartboard}, emcee \citep{foreman-mackey13}, F2PY \citep{f2py}, matplotlib \citep{matplotlib}, numpy \citep{numpy}, scipy \citep{scipy}}

\bibliographystyle{aasjournal}
\bibliography{references}

\appendix

\section{Walker initialization, Burn-in, and Autocorrelation} 
\label{sec:initialize}

Typically, only a small portion of the multi-dimensional parameter space has a non-zero posterior probability, and it is necessary to initialize the walkers in this small region. We initialize our walkers using a multi-step procedure. For each of our walkers, we randomly test positions in the parameter space until we find one with a non-zero posterior probability. From experience, we have found that many of these positions may be in very low probability regions. These walkers may become stuck and the sampler may not find a better position for many tens of thousands of steps, despite its low posterior probability.

After each walker has been placed at a random, non-zero position in parameter space, we then select the walker with the highest posterior probability. Following the advice of \citet{foreman-mackey13}, we move each of the walkers to a multi-dimensional ``ball'' in a region of high posterior probability; each walker is placed in a randomly selected position drawn from a narrowly peaked Gaussian centered on the best walker, ensuring the posterior probability is non-zero. We find that, from this initialization, the ensemble algorithm is generally efficient at expanding into the surrounding region of viable parameter space.

\begin{figure*}
\begin{center}
\includegraphics[width=0.8\textwidth]{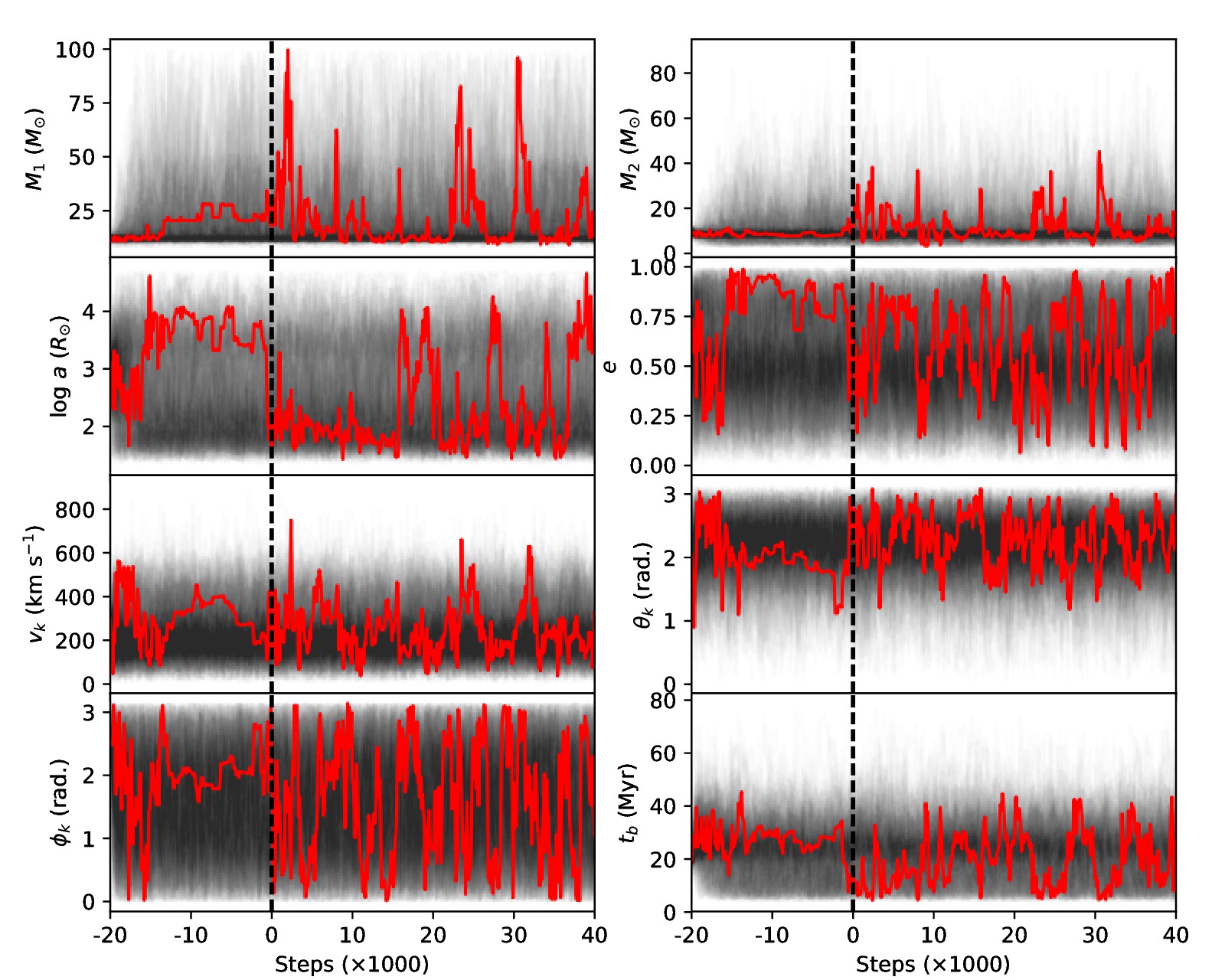}
\caption{The individual panels show the trace of the first 60,000 walker positions for each of the eight model parameters in our model of HMXBs described in Section \ref{sec:results_population}. For all models, we remove the first 20,000 steps as the burn-in, the points to the left of the dashed, vertical line. However, typical burn-in times vary substantially between models, and for this model, the trace indicates that a shorter burn-in is probably sufficient. One randomly chosen walker is highlighted in red, demonstrating that the walkers are able to move around the parameter space and typically do not become stuck in local minima. }
\label{fig:HMXB_chains}
\end{center}
\end{figure*}

Once the walkers have been set, we begin the burn-in stage. For the populations described in this work, we burn-in the model for 20,000 steps so the walkers distribute around the parameter space based on the posterior probability. We find this is a sufficient number for the chains to converge based on visual inspection of the trace of the chains. Figure \ref{fig:HMXB_chains} shows the trace of our 320 walkers for the first 40,000 steps for our HMXB model described in Section \ref{sec:results_population}. Although there is no formal proof that the walkers have completely converged, there is a difference in the evolution of the walkers during the burn-in and later in the evolution of the chains.

After our burn-in for 20,000 steps, we run our production models for 200,000 steps. Conclusions are only drawn from the production stage; data from the burn-in stages are removed. It is clear that the distribution of walkers in Figure \ref{fig:HMXB_chains} is not entirely smooth, a result of the complex posterior probability over these dimensions (for instance, multiple evolutionary channels). This is most obvious from the distribution in $a$, which has an overdensity at short separations (as discussed in Section \ref{sec:results_population}). One randomly selected chain is colored red to demonstrate the typical evolution of an individual walker. The highlighted walker demonstrates that walkers can move throughout the parameter space between different evolutionary channels.

\begin{figure}
\begin{center}
\includegraphics[width=0.45\columnwidth]{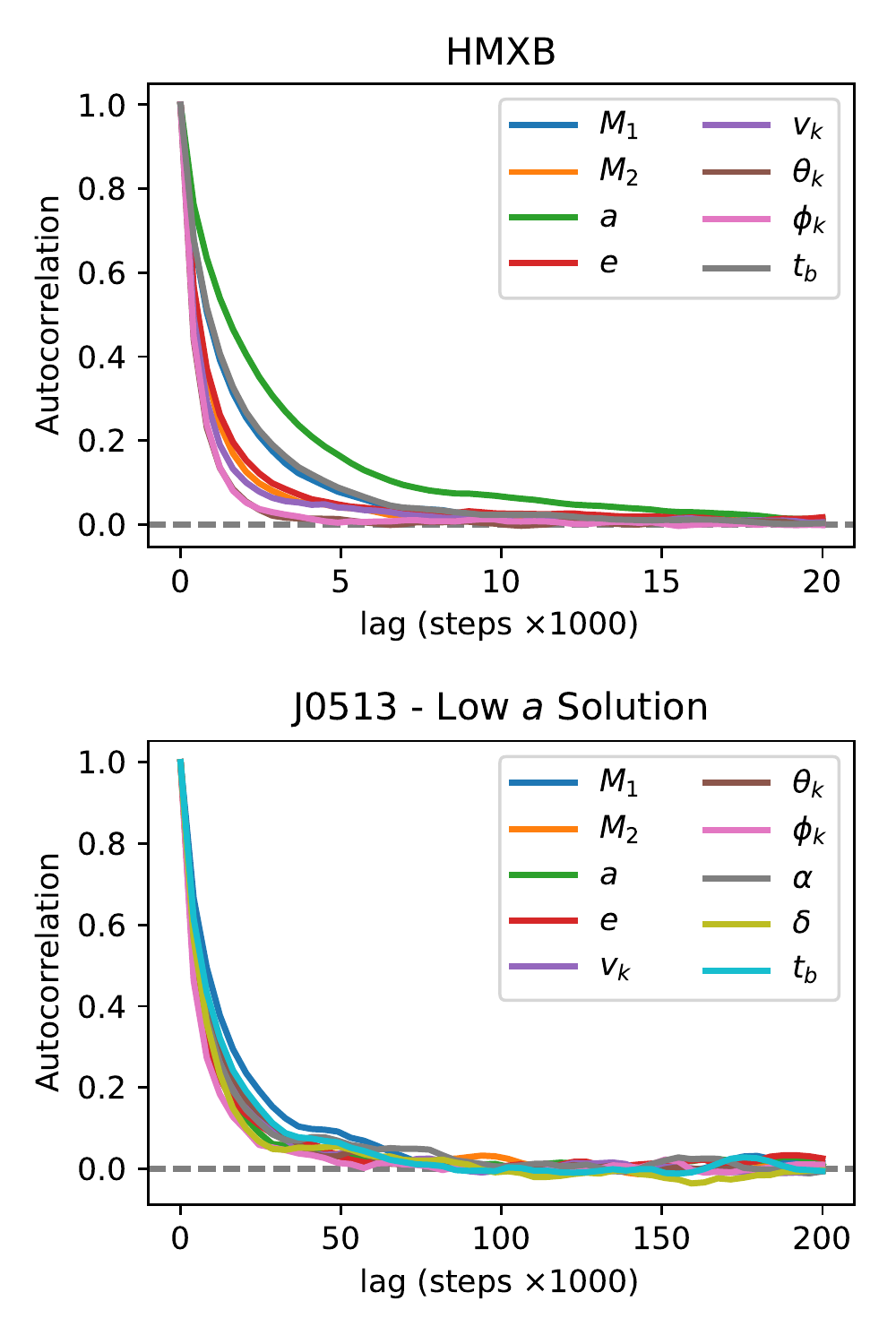}
\caption{The autocorrelation function for each of the parameters for our model of the population of HMXBs (top panel; as described in Section \ref{sec:results_population}) and of J0513 specifically (bottom panel; as described in Section \ref{sec:J0513_model}). These are calculated on the combined chains, in which each chain is stacked end-to-end for each parameter. Typical autocorrelation lengths for our eight parameter model are 5000-10,000. For a model in which the position and SFHs are included, autocorrelation lengths can be as much as a factor of 10 longer. In the case of J0513, the autocorrelation length is nearly 100,000 steps. }
\label{fig:HMXB_acor}
\end{center}
\end{figure}

Quantitatively, we can estimate how easily an individual walker can move around the parameter space using an autocorrelation function \citep[see discussion in][]{foreman-mackey13}. This determines how many steps need to be simulated before a walker has reached an effectively independent position in parameter space. Since our MCMC sampler uses an ensemble of 320 walkers, we stack them end-to-end to produce one long chain. Figure \ref{fig:HMXB_acor} shows the autocorrelation function for the long chain from the eight parameters in our HMXB model described in Section \ref{sec:results_population} in the top panel and for our ten parameter model of J0513 described in Section \ref{sec:J0513_model} in the bottom panel. The autocorrelation length is defined by the longest length of any individual parameter, in this case, $a$. For this model, we find an autocorrelation length of 5000-10,000 steps. The bottom panel shows that our model for J0513 has a substantially longer autocorrelation length of nearly 100,000 steps. Autocorrelation lengths must be calculated separately for each model, and clearly these differ significantly depending on the particular model run. One should be wary that enough steps need to be run before robust results are derived.

\section{{\tt BSE} in {\tt Python}} 
\label{sec:pybse}

We alter the publicly available stellar evolution code \bse\ to run within {\tt python}, which requires several adjustments. First, we create a wrapper around the main stellar evolution function call to include various model parameters (e.g., the common envelope efficiency) as inputs. The updated version of \bse\ is then compiled into a python module using {\tt f2py}. We then create a second, higher-level python module which calls this module. We include a {\tt setup.py} file for a user to add to their python libraries. This version of \bse\ is provided along with \dart.

With the exception of the reparameterization of supernova kicks, as discussed below in Section \ref{sec:supernova}, and the required adjustments to the function calls, we have not adjusted the physics within \bse. This approach allows for easy adjustment by different groups to import their own updates to prescriptions within \bse.

\section{Supernova Kicks}
\label{sec:supernova}

As mentioned in Section \ref{sec:intro}, our approach is to treat the supernova kick parameters as model parameters rather than random variables to be chosen within the stellar evolution code. While this requires editing of the function calls within \bse\ so that the variables may be propagated throughout the code, this is a necessity using our approach.

We calculate the post-SN orbital separation, $a_{\rm SN}$, systemic velocity, $v_{\rm sys}$, and eccentricity, $e_{\rm SN}$, based on the equations in \citet{hills83} and \citet{kalogera96}. In particular, we follow the reference frame from \citet{kalogera96} which differs from the default described by \citet{hurley02}. Our reference frame (at the instant of SN) sets the collapsing object at the origin, but with the velocity of the companion, such that the primary is moving with orbital velocity, $v_{\rm r}$. The supernova kick is parameterized by a magnitude, $v_{\rm k}$, a polar angle, $\theta_{\rm k}$, and an azimuthal angle, $\phi_{\rm k}$. $\theta_{\rm k}$ defines the angle between the kick velocity and the direction of orbital motion. This reference frame is optimal since, when eccentricity is not an observable this parameterization effectively removes a parameter.

We start by determining $a_{\rm SN}$ based on energy conservation:
\begin{equation}
a_{\rm SN} = \left[ \frac{2 }{a_{\rm pre-SN}}  - \frac{v_1^2}{G(M_{\rm CO} + M_2)} \right]^{-1}, \label{eq:SN_A} \\
\end{equation}
where $a_{\rm pre-SN}$ is the pre-SN orbital separation, $M_{\rm CO}$ is the post-SN compact object mass, $M_2$ is the companion mass, and $v_1$ is the post-kick velocity of the primary (in the reference frame of an initially stationary secondary):
\begin{equation}
v_1^2 = 2v_{\rm k} v_{\rm orb} \cos \theta_{\rm k} + v_{\rm k}^2 + v_{\rm orb}^2. \label{eq:v_1}
\end{equation}
The pre-SN orbital velocity, $v_{\rm orb}$ is defined as:
\begin{equation}
v_{\rm orb} = \sqrt{\frac{G (M_{\rm 1, pre-SN} + M_2)}{a_{\rm pre-SN}}}. \label{eq:v_r}
\end{equation}
The post-SN systemic velocity is:
\begin{equation}
v_{\rm sys}^2 = \beta^2_{\rm SN} v_{\rm k}^2
   + v_{\rm orb}^2 \left( \alpha_{\rm SN} - \beta_{\rm SN} \right)^2
   + 2 \beta_{\rm SN} v_{\rm orb} v_{\rm k} \cos \theta_{\rm k} \left( \alpha_{\rm SN} - \beta_{\rm SN} \right)
    \label{eq:SN_v_sys}
\end{equation}
where we have included two substitutions:
\begin{eqnarray}
\alpha_{\rm SN} &=& \frac{M_{\rm 1, pre-SN}}{M_{\rm 1, pre-SN} + M_2} \\
\beta_{\rm SN} &=& \frac{M_{\rm CO}}{M_{\rm CO} + M_2}
\end{eqnarray}

The post-SN eccentricity is determined by angular momentum conservation:
\begin{equation}
1-e_{\rm SN}^2 = \frac{a_{\rm pre-SN}^2}{a_{\rm SN}\ G (M_{\rm CO} + M_2)}  \left( v_{\rm k}^2 \cos^2\theta_{\rm k} + v_{\rm k}^2 \sin^2 \theta_{\rm k} \sin^2 \phi_{\rm k} + 2 v_{\rm k} v_{\rm orb} \cos \theta_{\rm k} + v_{\rm orb}^2  \right). \label{eq:SN_e}
\end{equation}
Systems with $0 \leq e < 1$ remain bound. We see that $\phi_{\rm k}$ only affects the post-SN orbit through the $\sin^2 \phi_{\rm k}$ term when solving for the post-SN orbital eccentricity in Equation \ref{eq:SN_e}.

It should be noted that this approach is only appropriate for systems which are circularized prior to undergoing core collapse, an assumption satisfied by the vast majority of binary systems \citep[eccentricity induced by the long term dynamical interaction with an external third body is one exception;][]{pijloo12}. Eccentric binaries could be taken into account by including the mean anomaly at the time of core collapse as an additional model parameter. We leave such a model for interested users as only rare evolutionary channels are eccentric at the time of core collapse.

\section{Consistency Checks and Benchmarking}
\label{sec:benchmarking}

In Section \ref{sec:mock}, we use \dart\ to recover the input parameters of three mock systems. These provide an important check of the method, however three mock systems do not form a complete test. We further want to make sure that by adding the position of the binary as an observable, a central advantage of using \dart\ over traditional population synthesis, the results for the other binary parameters are unbiased.

\begin{figure*}
\begin{center}
\includegraphics[width=0.95\textwidth]{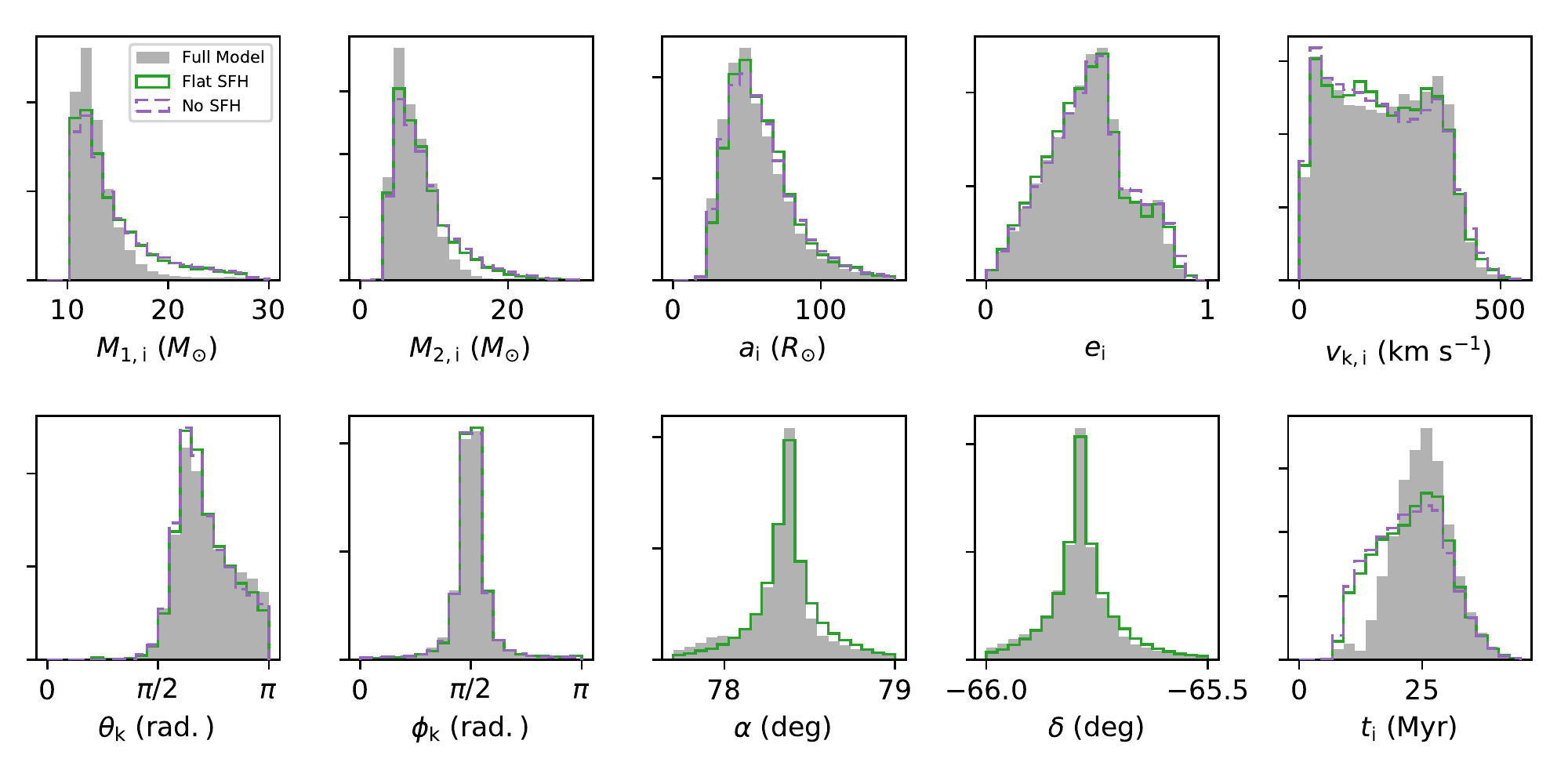}
\caption{ The normalized, 1D histograms for the ten parameters in our model for J0513 initialized with a short $a_{\rm i}$. We compare the histograms from three different models: our standard model (``Full Model''; filled, grey) as presented in Section \ref{sec:J0513_model}, one with a flat SFH throughout the LMC as described in Section \ref{sec:including_SFH} (``Flat SFH''; solid, green) , and an eight parameter model in which the position is ignored (the star formation rate is flat in this case) as described in Appendix \ref{sec:benchmarking} (``No SFH''; dashed, purple). The close similarity between the 1D histograms of ``Flat SFH'' and ``No SFH'' models indicates that our model passes an important consistency check: both models ought to produce identical results, as they both have a constant star formation rate.}
\label{fig:J0513_x_i_low}
\end{center}
\end{figure*}

\begin{figure*}
\begin{center}
\includegraphics[width=0.95\textwidth]{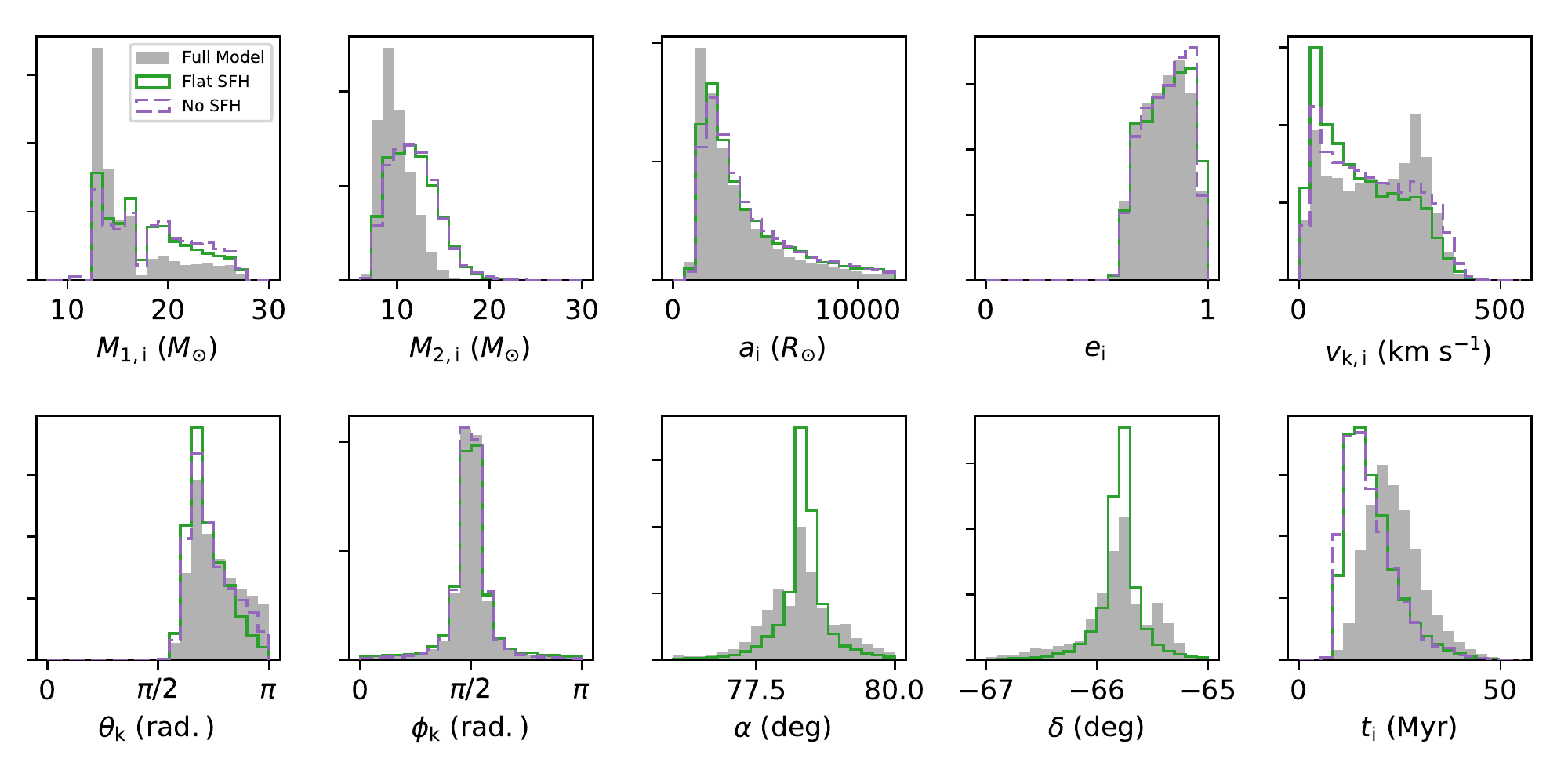}
\caption{ The normalized, 1D histograms for the ten parameters in our model for J0513 initialized with a long $a_{\rm i}$. The three models are the same as in Figure \ref{fig:J0513_x_i_low}. As expected, the ``Flat SFH'' and ``No SFH'' models produce essentially identical 1D distributions. }
\label{fig:J0513_x_i_high}
\end{center}
\end{figure*}





We perform a simple consistency check to ensure that the inclusion of birth position does not incorrectly bias our results. In our model of J0513 described in Section \ref{sec:J0513_model}, we use a flat SFH for the LMC; the birth right ascension and declination are therefore nuisance parameters, since no extra information has been added by including the star formation as a prior. Indeed, as we show in Figure \ref{fig:J0513_flatsfh_corner}, the posterior distribution in the birth position forms an axisymmetric distribution around its current position. We additionally generate a new model of J0513 using eight parameters, ignoring position information, but still using the other three observables provided in Table \ref{tab:J0513} as constraints on the system. The prior on $t_{\rm i}$ is flat in this model (i.e., the star formation rate is constant). For a correct implementation of the math described in Section \ref{sec:stats}, we should obtain identical posterior distributions for these eight parameters as in our model with a flat SFH for the LMC. For each of these additional models, we run two separate models: one initialized with a short $a_{\rm i}$ and one initialized with a long $a_{\rm i}$.

\begin{figure*}
\begin{center}
\includegraphics[width=0.95\textwidth]{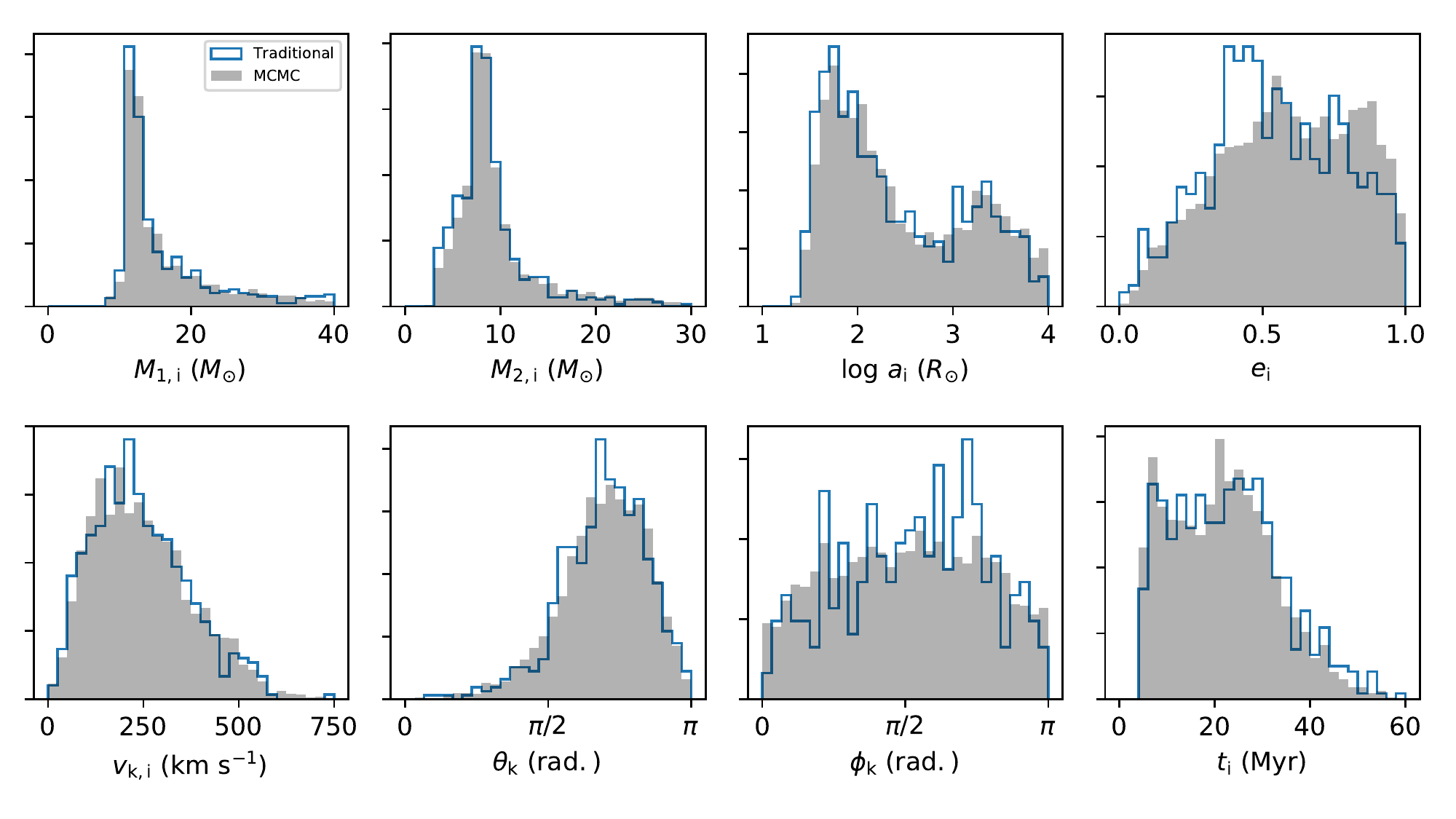}
\caption{ The normalized, 1D histograms for the eight parameters in our model for HMXBs described in Section \ref{sec:results_population} (filled grey). Blue distributions show the results for a sample of HMXBs using the same BPS model, but produced using traditional sampling methods. Although the two distributions were run for the same computation time, \dart\ produces a smoother posterior distribution. }
\label{fig:HMXB_x_i}
\end{center}
\end{figure*}

We compare the posterior distributions of model parameters in Figure \ref{fig:J0513_x_i_low} for the three different models of J0513 initialized with a short $a_{\rm i}$: our standard model (``Full Model''; filled, grey) as presented in Section \ref{sec:J0513_model}, one with a flat SFH throughout the LMC as described in Section \ref{sec:including_SFH} (``Flat SFH''; solid, green), and an eight parameter model in which the position is ignored (``No SFH''; dashed, purple). For a clear comparison we only show the 1D distributions of initial parameters for the Figure \ref{fig:J0513_x_i_low}. We provide the analogous comparison between these three models when initialized with a long $a_{\rm i}$ in Figure \ref{fig:J0513_x_i_high}. In both figures, the posterior distributions are nearly identical for our ``Flat SFH'' model and our ``No SFH'' model, as desired. Quantitatively, the similarity between the distributions holds up: the right two columns for each of the short $a_{\rm i}$ and long $a_{\rm i}$ solutions in Table \ref{tab:J0513_results} show that the median and 1$\sigma$ confidence levels for our ``Flat SFH'' model and our ``No SFH'' model are nearly identical. The model for J0513, using the actual SFH for the LMC differs significantly in Figures \ref{fig:J0513_x_i_low} and \ref{fig:J0513_x_i_high}, as expected since we demonstrate in Section \ref{sec:including_SFH} that using the correct SFH adds important constraints on the model.

As a second consistency check, we compare the posterior distribution of model parameters against the distribution produced by BPS using traditional sampling methods. Figure \ref{fig:HMXB_x_i} compares the 1D distributions from our MCMC approach using \dart\ for the HMXB model described in Section \ref{sec:results_population} with the results from BPS using traditional sampling methods. As done previously, we only show the 1D distributions for clarity. Figure \ref{fig:HMXB_x_i} shows that the two methods reach the same result. The only difference is that the traditional method produces a somewhat coarser distribution; we have checked with longer simulations that the two distributions indeed converge to the same result.

Why did we not run our simulation for more trials? To produce the sample using traditional methods, we generated random samples from the prior distributions, using the same amount of computation time as was used by \dart\ to produce the MCMC sample in Figure \ref{fig:HMXB_x_i}. In this case at least, the samples produced using our MCMC approach generate a smoother distribution, demonstrating the efficiency of this new approach.

\section{Bayesian Evidence Ratios}
\label{sec:evidence}

When multiple evolutionary channels can viably form an observed system, it is extremely helpful when the MCMC algorithm employed can efficiently explore the entire parameter space to identify the global maximum in posterior probability space. This was the case for our three mock systems; however when applying our model to J0513, the MCMC walkers were confined within one of two separate evolutionary channels. As discussed previously, in such cases we suggest separate simulations be run in which the each simulations' walkers be initialized near their respective local maxima. The simulations then separately explore the posterior space around each local maxima.

How then to determine which evolutionary channel is statistically preferred? Mathematically, for two different evolutionary channels, $C_1$ and $C_2$, we would like to determine the ratio between $P(C_1 \given \bm{D})$ and $P(C_2 \given \bm{D})$. Using Bayes' theorem, we can convert this ratio into a ratio between the products of a likelihood and a prior:
\begin{equation}
    \frac{P(C_1 \given \bm{D})}{P(C_2 \given \bm{D})} = \frac{P(\bm{D} \given C_1)}{P(\bm{D} \given C_2)} \frac{P(C_1)}{P(C_2)}, \label{eq:evidence_1}
\end{equation}
where we have canceled out terms representing the Bayesian evidence, $P(\bm{D})$, from both the numerator and denominator. The first term on the right hand side of Equation \ref{eq:evidence_1} is a ratio of quantities that are similar to the Bayesian evidence, but are conditional on the individual evolutionary channels. We will call these the Bayesian ``channel evidence." The ratio of the channel evidences is analogous to a Bayes factor. The second term on the right hand side is simply the branching ratio between the two evolutionary channels as dictated by the model, irrespective the system-specific constraints.

We next marginalize over the entire $\bm{x}_{\rm i}$ model parameter space:
\begin{equation}
    \frac{P(C_1 \given \bm{D})}{P(C_2 \given \bm{D})} = \frac{\int \dd \bm{x}_{\rm i}\ P(\bm{D} \given \bm{x}_{\rm i}, C_1) P(\bm{x}_{\rm i} \given C_1)} 
    {\int \dd \bm{x}_{\rm i} P(\bm{D} \given \bm{x}_{\rm i}, C_2) P(\bm{x}_{\rm i} \given C_2)}
    \frac{P(C_1)}{P(C_2)}. \label{eq:evidence_2}
\end{equation}
We can approximate these integrals using Monte Carlo sums:
\begin{equation}
    \frac{P(C_1 \given \bm{D})}{P(C_2 \given \bm{D})} \approx \frac{N_1^{-1}\sum_k^{N_1} P(\bm{D} \given \bm{x}_{{\rm i},k}, C_1) } 
    {N_2^{-1}\sum_l^{N_2} P(\bm{D} \given \bm{x}_{{\rm i},l}, C_2) }
    \frac{P(C_1)}{P(C_2)}, \label{eq:evidence_3}
\end{equation}
where $\bm{x}_{{\rm i},k}$ and $\bm{x}_{{\rm i},l}$ are random variates drawn from the following distributions:
\begin{eqnarray}
    \bm{x}_{{\rm i},k} &\sim& P(\bm{x}_{\rm i} \given C_1) \\
    \bm{x}_{{\rm i},l} &\sim& P(\bm{x}_{\rm i} \given C_2).
\end{eqnarray}

There is freedom over the choice of the number of samples we randomly draw in Equation \ref{eq:evidence_3}; we only require that both $N_1$ and $N_2$ are large enough that the summations approximate the integrals with sufficient accuracy. We can therefore choose $N_1$ and $N_2$ such that the ratio between them is exactly equal to the branching ratio of the two channels. Equation \ref{eq:evidence_3} then reduces to:
\begin{equation}
    \frac{P(C_1 \given \bm{D})}{P(C_2 \given \bm{D})} \approx \frac{\sum_k^{N_1} P(\bm{D} \given \bm{x}_{{\rm i},k}, C_1) } 
    {\sum_l^{N_2} P(\bm{D} \given \bm{x}_{{\rm i},l}, C_2) }. \label{eq:evidence_4}
\end{equation}

To evaluate these summations, note that $\bm{x}_{{\rm i},k}$ and $\bm{x}_{{\rm i},l}$ are randomly drawn from distributions without any dependence on $\bm{D}$. To obtain these samples, we run a secondary simulation, with identical prior distributions (including SFH), model parameters, and indicator function, but without any contribution to the likelihood function from system-specific observations. For J0513, we need not necessarily run an additional simulation, as we can use the results from our simulation of HMXBs in the LMC described in Section \ref{sec:results_population_LMC}. Unfortunately, there are not enough posterior samples with non-neglible likelihood to accurately calculate the ratio between the channel evidences with these simulations. Instead, we rerun this simulation, identical to our previous one, except we run it for 600,000 steps, and we add an additional constraint confining the birth position to be within 2\degree\ of the location of J0513.

Posterior samples from this secondary run are split into those going through $C_1$ and $C_2$, and {\it all} samples are used as random draws of $\bm{x}_{{\rm i},k}$ and $\bm{x}_{{\rm i},l}$ to calculate the sums in Equation \ref{eq:evidence_4}. Since the branching ratio $P(C_1)/P(C_2)$ is equal to the ratio of the number of posterior samples evolving through each channel, no extra calculation is necessary to determine the numbers of samples, $N_1$ and $N_2$, to draw from the posterior distribution for each channel. According to Equation \ref{eq:evidence_4}, ratios larger than unity indicate a preference for channel $C_1$, whereas those with ratios less than unity prefer channel $C_2$.

\begin{figure*}
\begin{center}
\includegraphics[width=0.5\textwidth]{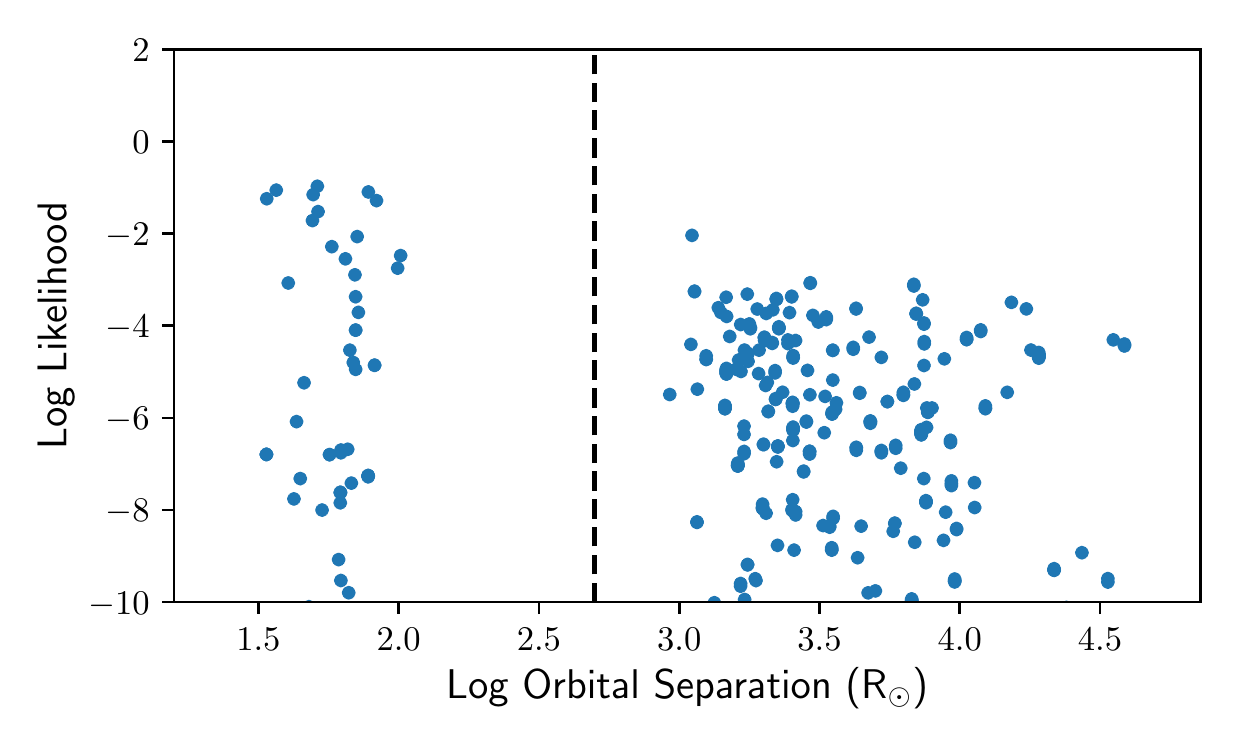}
\caption{ The highest likelihood posterior samples from our additional run as described in Section \ref{sec:evidence}, to calculate the channel evidence ratio for J0513. The samples show two clear clusters corresponding to the short $a_{\rm i}$ and long $a_{\rm i}$ solutions. To determine the preferred channel, the ratio of the channel evidences must be calculated. }
\label{fig:J0513_samples}
\end{center}
\end{figure*}

We show the highest likelihood posterior samples from our additional run in Figure \ref{fig:J0513_samples}. The two sets of samples neatly cluster in orbital separation space; from this figure it is apparent that the short $a_{\rm i}$ solution produces points with a larger likelihood, but the long $a_{\rm i}$ solution may cover a broader range in $a_{\rm i}$. To determine the preferred channel, the ratio of the channel evidences must be calculated. 

Given the dynamic range of the likelihoods shown in Figure \ref{fig:J0513_samples}, it is clear that only a handful of posterior samples dominate the channel evidence ratio calculation. Yet, we can still make important conclusions from these, since we only require enough samples with non-negligible likelihoods to determine which channel is preferred. Once determined, the appropriate posterior samples from our system-specific simulations provide the topology of the posterior space. 

\begin{figure*}
\begin{center}
\includegraphics[width=0.5\textwidth]{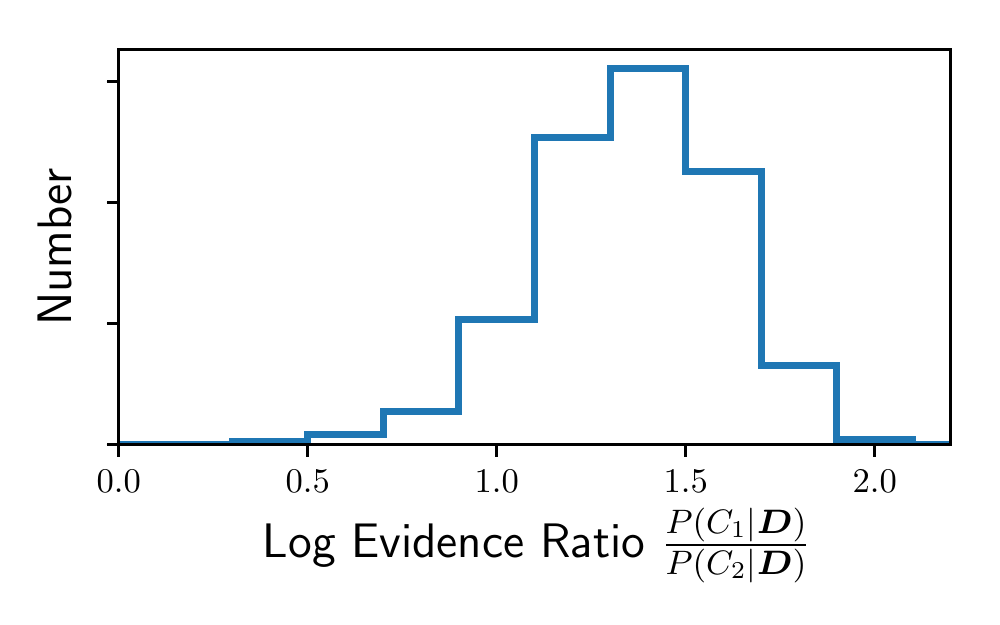}
\caption{ The distribution of logarithm of ratio of the channel evidences as calculated using a bootstrap method. Positive values indicate the ``short'' channel is preferred. Our bootstrap method suggests that the ``short" channel is preferred by a factor of 10-50. }
\label{fig:J0513_evidence}
\end{center}
\end{figure*}

Since we are dependent on only a few samples, we use a bootstrap method to determine the uncertainty in our channel evidence ratio calculation. We perform 1000 draws of the posterior samples from our secondary simulation, leaving out a random 5\% of the samples. The resulting distribution of channel evidence ratios is shown in Figure \ref{fig:J0513_evidence}. This distribution indicates that the the short $a_{\rm i}$ solution is preferred over the long $a_{\rm i}$ solution by a factor of $\gtrsim$10.

When simulating J0513, the multiple evolutionary channels were easily separated in one parameter, the orbital separation. This need not be generally true, and two separate, local maxima may be separated in parameter space by some high-dimensional manifold that may be difficult to visualize or parameterize. In such cases, clustering and classification algorithms may provide data-driven approaches to identifying the evolutionary channels to which each posterior sample is a member.

In all our testing we found no more than two evolutionary channels (of course, many variations on these two channels exist, but the MCMC walkers were only confined to two broadly defined, separate regions, and then only when applied to J0513 with its precisely observed parameters). This need not always be the case. If a user is at all concerned about the possibility of the MCMC walkers missing the global maximum due to partitioning of the posterior space by deep valleys, we suggest secondary simulations be run with \dart\ using identical prior distributions, model parameters, and indicator function, but without any system-specific constraints.  We then suggest comparing the resulting distribution of posterior samples, weighted by each sample's likelihood (calculated using the system-specific constraints), to posterior samples from the full, system-specific run. Differences between the two distributions could be due to additional evolutionary channels missed by the MCMC walkers. Once identified, alternative evolutionary channels can be explored using new MCMC simulations initialized around high-likelihood samples from the secondary simulation.

Depending on the precision of the observations, this secondary run may take a very long time to produce enough samples to identify all the viable evolutionary channels. It is always possible that a rare solution exists with a very low prior probability, but very high likelihood, such that the evolutionary channel dominates the overall Bayesian evidence. If such a channel is isolated from other local maxima and no simulation is run with walkers initialized near this location in parameter space, the statistically preferred formation scenario may be missed altogether. It is worth noting that this is a problem that also afflicts traditional population synthesis methods; however note that because a grid-based method searches a broader range in parameter space, one can calculate the maximum contribution to the Bayesian evidence of any alternative solutions that fall in-between tested grid points.

\end{document}